\documentclass[a4paper,twocolumn,10pt,accepted=2023-03-12]{quantumarticle}
\pdfoutput=1
\usepackage[utf8]{inputenc}
\usepackage[english]{babel}
\usepackage[T1]{fontenc}
\usepackage{amsmath}
\usepackage{amsfonts}
\usepackage{amssymb}
\usepackage{hyperref}
\usepackage{faktor}
\usepackage{mathtools}
\usepackage{tikz}
\usepackage{lipsum}
\usepackage{braket}
\usepackage{listings}
\usepackage{cleveref}
\usepackage{amsthm}
\usepackage{float}
\theoremstyle{definition}
\newtheorem{example}{Example}[section]

\usepackage{caption}
\usepackage{subcaption}

\DeclareRobustCommand\openone{\leavevmode\hbox{\small1\normalsize\kern-.33em1}}
\DeclareRobustCommand\openonesmall{\leavevmode\hbox{\footnotesize1\small\kern-.30em1}}

\usepackage{algorithm,algpseudocode}
\algnewcommand{\LineComment}[1]{\Statex \Statex \(\triangleright\) #1}

\renewcommand{\arraystretch}{1.2}

\makeatletter
\newenvironment{breakablealgorithm}
{
	\begin{center}
		\refstepcounter{algorithm}
		\hrule height.8pt depth0pt \kern2pt
		\renewcommand{\caption}[2][\relax]{
			{\raggedright\textbf{\fname@algorithm~\thealgorithm} ##2\par}%
			\ifx\relax##1\relax 
			\addcontentsline{loa}{algorithm}{\protect\numberline{\thealgorithm}##2}%
			\else 
			\addcontentsline{loa}{algorithm}{\protect\numberline{\thealgorithm}##1}%
			\fi
			\kern2pt\hrule\kern2pt
		}
	}{
		\kern2pt\hrule\relax
	\end{center}
}

\usepackage[numbers,sort&compress]{natbib}

\newcommand{\joschka}[1]{\textcolor{black}{#1}}
\newcommand{\joschkaedit}[1]{\textcolor{black}{#1}}

\newcommand{\jr}[1]{\textcolor{black}{#1}} 
\newcommand{\etc}[1]{\textcolor{black}{#1}}

\begin{document}

\title{Bias-tailored quantum LDPC codes}

\author{Joschka Roffe}
\affiliation{Dahlem Center for Complex Quantum Systems, Freie Universität Berlin, 14195 Berlin, Germany}
\affiliation{Department of Physics and Astronomy, University of Sheffield, Sheffield S3 7RH, United Kingdom}
\email{joschka@roffe.eu}
\homepage{https://roffe.eu}
\orcid{0000-0001-9202-1156}
\author{Lawrence Z. Cohen}
\affiliation{Centre for Engineered Quantum Systems, School of Physics,
University of Sydney, Sydney, New South Wales 2006, Australia}
\author{Armanda O. Quintavalle}
\affiliation{Department of Physics and Astronomy, University of Sheffield, Sheffield S3 7RH, United Kingdom}
\affiliation{Riverlane, Cambridge  CB2 3BZ, United Kingdom}
\author{Daryus Chandra}
\affiliation{School of Electronics and Computer Science, University of Southampton, Southampton SO17 1BJ, United Kingdom}
\author{Earl T. Campbell}
\affiliation{Department of Physics and Astronomy, University of Sheffield, Sheffield S3 7RH, United Kingdom}
\affiliation{Riverlane, Cambridge  CB2 3BZ, United Kingdom}
\affiliation{AWS Center for Quantum Computing, Cambridge CB1 2GA, United Kingdom}

\maketitle

\begin{abstract}
\noindent Bias-tailoring allows quantum error correction codes to exploit qubit noise asymmetry. Recently, it was shown that a modified form of the surface code, the XZZX code, exhibits considerably improved performance under biased noise. In this work, we demonstrate that quantum low density parity check codes can be similarly bias-tailored. \jr{We introduce a bias-tailored lifted product code construction that provides the framework to expand bias-tailoring methods beyond the family of 2D topological codes. We present examples of bias-tailored lifted product codes based on classical quasi-cyclic codes and numerically assess their performance using a belief propagation plus ordered statistics decoder.} Our Monte Carlo simulations, performed under asymmetric noise, show that bias-tailored codes achieve several orders of magnitude improvement in their error suppression relative to depolarising noise.
\end{abstract}

\section{Introduction}

All current quantum computing architectures are subject to the same fundamental problem: quantum bits (qubits) are extremely susceptible to error. To counter this, quantum error correction (QEC) \cite{Shor1995,roffe2019quantum} must be incorporated into the quantum computing stack. With an appropriate choice of QEC code, it is in principle possible to build arbitrarily large fault tolerant quantum computers. However, the trade-off is that QEC can considerably increase the device overhead in terms of the total number of qubits required.  

Quantum codes are commonly designed and benchmarked under the assumption that their qubits are subject to depolarising noise for which $X$ (bit), $Z$ (phase) and $Y$-type Pauli errors occur with equal probability. In practice, however, physical qubits are subject to biased noise where one species of error is stronger than the others. For example, in some superconducting qubit architectures phase-noise dominates by as much as five orders of magnitude \cite{aliferis2009,lescanne,aws_architecture}. A \textit{bias-tailored} quantum code is designed to exploit knowledge of such noise asymmetry to improve QEC performance. Previous studies have shown that bias-tailored QEC codes, implemented in conjunction with bias-preserving two-qubit gates \cite{puri2020}, could considerably reduce the overhead associated with achieving fault tolerant quantum computation.

In recent work by Bonilla Ataides et al. \cite{ataides2020}, it was shown that a variant of the surface code -- the XZZX code -- can achieve remarkable performance under biased noise. The XZZX surface code is obtained from the standard surface code via a local modification of stabilisers \cite{wen2002,wootton2012,kovalev12}. The principal advantage of the XZZX surface code is that its structure simplifies to a set of decoupled repetition codes in the infinite-bias regime. As such, the XZZX surface code is readily decodable with a threshold upper-bounded at $50\%$ for infinite-bias. This infinite-bias threshold considerably improves on the corresponding threshold of $10.9\%$ \cite{bombin2012strong} for the standard surface code.

The surface code can be embedded onto a geometrically local architecture where gates are only performed between nearest-neighbours. This makes the surface code (and its variants) an attractive QEC scheme for the current generation of quantum architectures which typically have constrained inter-qubit connectivity \cite{takita2017,Arute2019,aws_architecture}. Despite this, the surface code may not be a practical solution for fault tolerance in full-scale quantum computers. The main drawback is that it has poor rate: each surface code patch encodes only a single logical qubit. As such, quantum computers based on the surface code will require high-qubit overheads \cite{gidney_rsa_2021}.

\jr{
Several no-go results suggest the construction of local (or near-local) high-rate, high-distance QEC codes will not be feasible \cite{bravyi2010tradeoffs,baspin2021connectivity,delfosse2021bounds}. Fortunately, various technologies are currently in development that promise long-range inter-qubit connectivity \cite{Debnath16, bergeron2020, Magnard2020, ramette2021}. On these platforms it will be possible to run \textit{quantum low density parity check (LDPC) codes} in place of the surface code \cite{breuckmann2021,cohen21}. Such quantum LDPC codes encode multiple qubits per logical block and promise a route to efficient fault tolerant quantum computation.}

\jr{
In classical communication, LDPC codes are some of the best performing error correction protocols, underpinning technologies such as the 5G mobile standard, Gigabit Ethernet and WiFi (802.11n, 802.11ac)~\cite{shao2019survey, tzimpragos2014survey}. At first, it was not known whether it would be possible to construct quantum LDPC codes with parameter scaling comparable to the best-known classical protocols. However, a series of recent advances have shown that asymptotically \textit{good} quantum LDPC do exist \cite{hastings2021fiber,breuckmann21,panteleev2020quantum}. Most recently, Pantaleev and Kalachev introduced \textit{lifted product codes} as a family of quantum LDPC codes with finite-rate and linear block-length-to-distance scaling \cite{panteleev2020quantum,panteleev21}. With the underlying theoretical foundation now set, the challenge remains in engineering quantum LDPC codes into practical architectures for scalable fault tolerant quantum computing.}

\subsection{Summary of results}

\jr{
In this paper, we introduce a \textit{bias-tailored lifted product} that enables the construction of quantum LDPC codes that exploit qubit noise asymmetry. Our bias-tailored lifted product is obtained via a local Clifford transformation of the stabilisers of the original lifted product introduced by Pantaleev and Kalachev \cite{panteleev2020quantum}. The new stabilisers ensure that in the limit of infinite-bias, the effective distance of the quantum code is increased. This enables bias-tailored lifted product codes to harness the increased code capacity of biased error channels. }

\jr{
We construct explicit examples of bias-tailored lifted product codes based on classical quasi-cyclic codes \cite{fossorier2004quasicyclic}. These codes have complex syndrome patterns meaning they cannot be decoded using the minimum-weight perfect matching algorithms that work well for topological codes. Instead, we show that bias-tailored lifted product codes can be efficiently decoded using a belief propagation plus ordered statistics decoder (BP+OSD) \cite{Panteleev_2019,roffe2020quantum}. The decoding routine uses a method that improves performance by exploiting knowledge of correlations that arise between the bit- and phase-decoding components due to Pauli-$Y$ errors. This procedure is outlined in detail in Appendix~\ref{app:simulation}, and our simulation scripts are available open source on Github \cite{biased_qldpc_code}. }

\jr{
To benchmark bias-tailored lifted product codes, we perform extensive BP+OSD decoding simulations across a number of bias regimes and compare to the case of depolarising noise. Our results show that the bias-tailored lifted product achieves improved logical error suppression for any biased error channel in which $X$- or $Z$-Pauli noise dominates. In the limit of infinite-bias, we observe logical error rates several orders of magnitude less than that obtained for depolarising noise. In contrast, quantum LDPC codes constructed from the original lifted product typically suffer from a reduction in code performance with increasing noise asymmetry. We also demonstrate that quantum codes can be specifically designed to respond better to $Z$-bias compared to $X$-bias (or vice-versa).}

\jr{We conjecture that the bias-tailored lifted product provides a versatile framework for the construction of codes that respond well to noise asymmetry. As an example of the construction's utility, we demonstrate that an XZZX toric code can be directly derived from the bias-tailored lifted product. Similar to the XZZX surface code, the XZZX toric code responds well to noise-asymmetry due to the fact it decouples to a set of independent repetition codes in the infinite-bias limit. Additionally, the XZZX toric code benefits from twisted periodic boundary conditions that lead to logical operators that wrap around the lattice. We show that both of these features arise automatically from the bias-tailored lifted product of two repetition codes. This demonstrates that the bias-tailored lifted product applies not just to the construction of random quantum LDPC codes, but also to structured topological codes.}

This paper is structured as follows. In Section~\ref{sec:ecc}, we cover background in classical coding theory. This includes a description of how quasi-cyclic LDPC codes are constructed from base protographs. Section~\ref{sec:qecc} outlines the essential elements of the stabiliser framework for QEC relevant to this work. In particular, we describe how quantum LDPC codes can be obtained from classical codes via product constructions. In Section~\ref{sec:bias_tailored_qec}, we present our methods for bias-tailoring QEC codes. We begin by unpicking the essential characteristics that enable XZZX codes to respond well to biased error models. We then show that the XZZX toric code can be described as a special case of a lifted product we refer to as the bias-tailored lifted product. This provides the framework through which the concept of bias tailoring can be generalised to quantum LDPC codes. Finally, in Section~\ref{sec:numerics}, we present the findings of our Monte Carlo decoding simulations of quantum LDPC codes under different bias regimes.

\section{Classical Error Correction} \label{sec:ecc}

\subsection{Binary codes}\label{sec:classical_error_correction}

A classical binary error correction code $\mathcal{C}$ describes a redundant mapping of a $k$-bit string $\textbf{b}$ to an $n$-bit codeword $\textbf{c}$, where $n>k$ \footnote{We use bold symbols to denote binary strings. For example, $\textbf{b}=01$. When involved in matrix operations we assume these binary strings are equivalent to column vectors. Eg. $\textbf{b}=01=\left[0,1\right]^T$.}. The codewords $\textbf{c}\in \mathcal{C}=\textsc{nullspace}(H)$ are the null-space vectors of an $m{\times} n$ binary parity check matrix $H$ such that $H\cdot\textbf{c} \ {\rm{mod \ 2}}=\textbf{0}$ (note that from this point onward, we will assume mod $2$ arithmetic for operations involving bit strings and parity check matrices). By the rank-nullity theorem, the number of bits $k$ encoded by code $\mathcal{C}$ is related to the rank of the parity check matrix $H$ such that $k=n-{\rm{rank}}(H)$. The code distance $d$ is defined as the minimum Hamming weight $d={\rm min}({\rm{wt}}(\textbf{c}))$ of a non-zero codeword $\textbf{c} \in \mathcal{C}$. If a codeword $\textbf{c} \in \mathcal{C} $ is subject to an error $\textbf{e}$, this error can be detected by computing the code syndrome $\textbf{s}=H\cdot (\textbf{c}+\textbf{e})=H\cdot\textbf{e}$. Code syndromes are non-zero for all errors $\textbf{e}$ of Hamming weight less than the code distance ${\rm wt}(\textbf{e})<d$. We adopt the convention of labelling classical codes in terms of the $[n,k,d]$ notation, where $n$ is the block length (codeword length), $k$ is the dimension (number of encoded bits) and $d$ is the code distance. Another important figure of merit is the code rate $r$, defined as the ratio of encoded bits to the block length $r=k/n$.

\begin{example}\label{example:rep_code}
The parity check matrix for a length-three closed-loop repetition code is given below
\begin{equation} \label{eq:parity_check_rep_code}
H^{\rm rep}_3 = \begin{bmatrix} 
    1 & 1 & 0 \\
    0 & 1 & 1 \\
    1 & 0 & 1 
    \end{bmatrix}\rm.\end{equation}
The codewords of $H^{\rm rep}_3$ are given by $\mathcal{C}={\rm \textsc{nullspace}}(H^{\rm rep}_3)=\{ 000,111\}$. The block length of this code is $n=3$. By the rank-nullity theorem, the number of encoded bits is $k=n-{\rm rank}(H)=1$. The lowest-weight non-zero codeword is $111$ giving a code distance of $d=3$. The $[n,k,d]$ code parameters of $H^{\rm rep}_3$ are therefore $[3,1,3]$. As an example of error detection, the error $\textbf{e}=100$ would give a syndrome $\textbf{s}=H \cdot \textbf{e} = 101$. The distance of a repetition code can be improved simply by increasing the length of the codewords (increasing the number of repetitions). In general, a length-$n$ repetition code will have  code parameters $[n,k,d] = [n,1,n]$.

\subsection{Factor graphs}
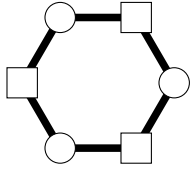
\begin{figure}
    \centering
    \begin{tikzpicture}
\draw[black,thick,line width=1.00mm] (1.0,0.0) -- (0.5000000000000001,0.8660254037844386);
\draw[black,thick,line width=1.00mm] (0.5000000000000001,0.8660254037844386) -- (-0.4999999999999998,0.8660254037844387);
\draw[black,thick,line width=1.00mm] (-0.4999999999999998,0.8660254037844387) -- (-1.0,1.2246467991473532e-16);
\draw[black,thick,line width=1.00mm] (-1.0,1.2246467991473532e-16) -- (-0.5000000000000004,-0.8660254037844384);
\draw[black,thick,line width=1.00mm] (-0.5000000000000004,-0.8660254037844384) -- (0.49999999999999933,-0.866025403784439);
\draw[black,thick,line width=1.00mm] (0.49999999999999933,-0.866025403784439) -- (1.0,0.0);
\filldraw[fill=white, draw=black] (1.000000,0.000000) circle (0.200000) node [label=center:{$\scriptstyle {} $}]{};
\filldraw[fill=white, draw=black] (0.300000,0.666025) rectangle (0.700000,1.066025) node [label=above right:{}]{};\filldraw[fill=white, draw=white] (0.500000,0.866025) circle (0.100000) node [label=center:{$\scriptstyle {} $}]{};
\filldraw[fill=white, draw=black] (-0.500000,0.866025) circle (0.200000) node [label=center:{$\scriptstyle {} $}]{};
\filldraw[fill=white, draw=black] (-1.200000,-0.200000) rectangle (-0.800000,0.200000) node [label=above right:{}]{};\filldraw[fill=white, draw=white] (-1.000000,0.000000) circle (0.100000) node [label=center:{$\scriptstyle {} $}]{};
\filldraw[fill=white, draw=black] (-0.500000,-0.866025) circle (0.200000) node [label=center:{$\scriptstyle {} $}]{};
\filldraw[fill=white, draw=black] (0.300000,-1.066025) rectangle (0.700000,-0.666025) node [label=above right:{}]{};\filldraw[fill=white, draw=white] (0.500000,-0.866025) circle (0.100000) node [label=center:{$\scriptstyle {} $}]{};
\end{tikzpicture}
    \caption{The factor graph of the distance-three closed loop repetition code. The circular nodes represent data bits whilst the square nodes represent parity bits. Edges are drawn between nodes according to the parity check matrix $H_3^{\rm rep}$ defined in Eq.~(\ref{eq:parity_check_rep_code}).}
    \label{fig:fg_rep_code}
\end{figure}
The factor graph of a code is a bipartite graph with an adjacency matrix given by the parity check matrix $H$: the columns of $H$ (data bits) map to the round nodes and the rows of $H$ (check bits) map to the square nodes. As an example, Fig.~\ref{fig:fg_rep_code}  shows a factor graph representation of the length-three repetition code. The corresponding adjacency matrix is given by $H_3^{\rm rep}$ defined in Eq.~(\ref{eq:parity_check_rep_code}). From the factor graph, we see that larger repetition codes can be obtained simply by increasing the length of the loop. In addition to providing a useful visualisation of the code, factor graphs serve a purpose in code design, analysis and decoding \cite{kschischang2001factor,roffe2020quantum}.
\end{example} 

\subsection{Quasi-cyclic LDPC codes}\label{sec:quasi_cyclic}

Low density parity check (LDPC) codes are classical binary codes with sparse factor graphs. More concisely, a code is labelled $(l,q)$-LDPC when the column and row weights of its parity check matrix are upper bounded by $l$ and $q$ respectively. There are many methods for constructing high-performance LDPC codes for a variety of use cases. In the setting of classical communication, one of the most widely used methods is the quasi-cyclic construction~\cite{fossorier2004quasicyclic}.

The fundamental building blocks of quasi-cyclic codes are circulant permutation matrices. We define a circulant permutation matrix $\lambda_L^t$ as the matrix obtained by shifting the columns of an $L{\times} L$ identity matrix $\openone_L$ by $t$ positions to the right. As an example, the $3{\times}3$ circulant permutation matrix with shift parameter $t=1$ is given by
\begin{equation}
\lambda^{1}_3=\begin{bmatrix}
0&1&0\\
0&0&1\\
1&0&0
\end{bmatrix}\rm.
\end{equation}
It is clear that after $L$ shifts, the circulant permutation matrix $\lambda^L_L$ is equal to the identity $\lambda^L_L=\lambda^0_L=\openone_L$. In general, circulant permutation matrices obey the cyclic property
\begin{equation}
\lambda^t_{L}=\lambda_L^{t \ {\rm mod} \ L}\rm.
\label{eq:cyclic_property}
\end{equation}
The set $\{ \lambda^0_L,\lambda^1_L,...,\lambda^{L-1}_L\}$ of $L$ distinct permutation matrices -- including the identity -- forms a basis of a vector space $\mathbb{A}_L$ over the binary field $\mathbb F_2$. Any element $\alpha \in \mathbb A_L$ can be written as a linear combination of the elements from this basis:
\begin{equation}\label{eq:ring_addition}
\alpha = \sum_{l = 0}^{L-1} a_l \lambda^l_L,
\end{equation}
where $a_l \in \mathbb{F}_2$. The product of any two circulant permutation matrices is also a permutation matrix, and it holds:
\begin{equation}
\lambda_L^i \cdot \lambda_L^j = \lambda_L^{i + j\ {\rm mod} \ L}\rm.
\end{equation}
In general, 
\begin{equation}\label{eq:ring_mult}
    \alpha\cdot\beta\in \mathbb{A}_L\rm,
\end{equation}
    for any pair of elements $\alpha,\beta \in \mathbb{A}_L$. Hence, $\mathbb{A}_L$ is a  \textit{ring} and we refer to it as the \textit{ring of circulants}\footnote{A ring $\mathbb A$ is a set equipped with addition and multiplication such that it is a group under addition and it is closed under multiplication. Since permutation matrices $\lambda_L^t$ satisfy the polynomial equation $x^L - 1 = 0$, $\mathbb{A}_L$ can be expressed as a quotient ring $\mathbb{F}_2[x]/(x^L-1)$. For more details, see, for instance,  \cite{childs2009concrete}.}.
    
The binary codes introduced in Section~\ref{sec:classical_error_correction} have parity check matrices with coefficients in $\mathbb{F}_2$. To construct a quasi-cyclic code, we first define an $m' {\times} n'$ matrix $A\in \mathbb{A}_L^{m'{\times} n'}$ with coefficients in the ring of circulants $\mathbb{A}_L$. This matrix is sometimes referred to as a \textit{protograph} and provides a compact means of representing a large quasi-cyclic code in terms of its constituent permutation matrices. We indicate by $\mathfrak{B}(.)$ the mapping that transforms a protograph to its binary representation
\begin{equation}\label{eq:binary_mapping}
    \mathfrak{B}(.): \quad A\in \mathbb{A}_L^{m'{\times} n'} \mapsto \mathfrak{B}(A)\in \mathbb{F}_2^{m\times n}\rm. 
\end{equation}
The binary matrix $\mathfrak{B}(A)$ is obtained from $A$ by replacing each coefficient $A_{ij} = \sum_{l = 0}^{L-1} \alpha_{l}\lambda_{L}^{l}$ of $A$ with its binary representation, i.e.\ by substituting each $\lambda_L^l$ with the corresponding permutation matrix in $\mathbb{F}_2^{L \times L}$. In general, an $m' {\times} n'$ protograph $A\in \mathbb{A}_L^{m'{\times} n'}$  yields an $m{\times} n$ binary matrix $\mathfrak{B}(A)\in \mathbb{F}_2^{m{\times} n}$ where $m=m' L$ and $n=n'L$.

We define the \emph{weight} of $\alpha \in \mathbb A_L$ as the number of non zero terms in the decomposition defined in Eq.~(\ref{eq:ring_addition}), i.e.\ if $\alpha = a_0 \openone + \dots + a_{L-1}\lambda_L^{L-1}$, ${\rm wt}(\alpha) = \lvert \{ i \text{ s.t } a_i = 1 \}\rvert$. For example, ${\rm wt}(\lambda^L_0)=1$ and ${\rm wt}(\lambda^0_L +\lambda^1_L)=2$. We associate to each protograph $A\in\mathbb{A}_L^{m'{\times} n'}$, a \emph{weight enumerator matrix} $W\in \mathbb{Z}^{m'{\times} n'}$ such that $W_{ij}={\rm wt}(A_{ij})$. Since each permutation matrix has rows and columns of weight one, it is easy to verify that the row/column weight of $W$ equates the row/column weight of the binary matrix $\mathfrak{B}(A)$. This equivalence allows us to bound the sparsity of the quasi-cyclic LDPC code with parity check matrix $\mathfrak{B}(A)$ in terms of the protograph $A$. A protograph whose weight enumerator matrix has row and column weights upper-bounded by $l$ and $q$ respectively yields an $(l,q)$-LDPC code.

\begin{example}
Consider the quasi-cyclic code defined by the following $2{\times}3$ protograph
\begin{equation}
A_L=
\begin{bmatrix}
\lambda^1_L+\lambda^2_L & \lambda^0_L & 0\\
0& \lambda^0_L+\lambda^1_L & \lambda^1_L
\end{bmatrix}
\end{equation}
The weight enumerator matrix of $A_L$ is given by
\begin{equation}
    W=\begin{bmatrix}
    2&1&0\\
    0&2&1
    \end{bmatrix}\rm.
\end{equation}
The column and row weights of $W$ are upper bounded by three and three respectively. Any parity check matrix $\mathfrak{B}(A_L)$ derived from the protograph $A_L$ will therefore be $(3,3)$-LDPC. When $L=3$, the protograph $A_3$ yields a $[9,3,3]$ code with the following parity check matrix
\begin{equation}
    \mathfrak{B}(A_3) =\left[\begin{array}{@{}ccc|ccc|ccc@{}}
    0&1&1&1&0&0&0&0&0\\ 
    1&0&1&0&1&0&0&0&0\\ 
    1&1&0&0&0&1&0&0&0\\ \hline
    0&0&0&1&1&0&0&1&0\\ 
    0&0&0&0&1&1&0&0&1\\ 
    0&0&0&1&0&1&1&0&0
    \end{array}\right]\rm.
\end{equation}
\end{example}

\begin{example} \label{ex:proto_rep}
The family of distance-$L$ closed-loop repetition codes can be interpreted as quasi-cyclic codes defined by the following $1{\times} 1$ protograph
\begin{equation}\label{eq:rep_proto}
A^{\rm rep}_L=\begin{bmatrix}
\lambda^0_L+\lambda^1_L
\end{bmatrix}\rm.
\end{equation}
The weight enumerator matrix of $A^{\rm rep}_L$ is $W=\left[2\right]$. As such, for any $L>1$, the repetition code with parity check matrix $\mathfrak{B}(A_L^{\rm rep})$ is $(2,2)$-LDPC. It is clear that, for $L = 3$, $\mathfrak{B}(A^{\rm rep}_3)$ is equal to the parity check matrix of the distance-three repetition code defined in Eq.~(\ref{eq:parity_check_rep_code}).  

\end{example}

\section{Quantum error correction}\label{sec:qecc}

\subsection{Stabiliser codes}

Similar to classical binary codes, a stabiliser QEC code describes a redundant mapping from a $K$-qubit state $\ket{\psi}$ to an $N$-qubit logical state $\ket{\psi}_L$. \jr{Qubit errors are described by linear combinations of matrices from the basis $\{\openone,X,Y,Z\}\in\mathcal{G}$, where $\mathcal{G}$ is the Pauli group}\footnote{The single-qubit Pauli group $\mathcal{G}_1=\langle I,X,Y,Z\rangle$ where $\openone=\left[\begin{smallmatrix}1&0\\0&1\end{smallmatrix}\right]$, $X=\left[\begin{smallmatrix}0&1\\1&0\end{smallmatrix}\right]$, $Y=\left[\begin{smallmatrix}0&-{\rm i}\\{\rm i}&0\end{smallmatrix}\right]$ and $Z=\left[\begin{smallmatrix}1&0\\0&-1\end{smallmatrix}\right]$. The Pauli-group over $N$-qubits is given by $\mathcal{G}_N=\bigotimes_{i=0}^N \mathcal{G}_i$.}. The logical states are defined so that they project onto the $(+1)$ eigenspace of a group of Pauli operators $\mathcal{S}\subset \mathcal{G}_N$ known as the code stabilisers such that $S\ket{\psi}_L=(+1)\ket{\psi}_L$ for all $S \in \mathcal{S}$. If the logical state is subject to a Pauli error $E$ that anti-commutes with a stabiliser $S\in \mathcal{S}$, then the measurement of this stabiliser projects onto the $(-1)$ eigenspace $S_jE \ket{\psi}_L=(-1)\ket{\psi}_L$. The results of  the stabiliser measurements form a syndrome that can be decoded to determine the best recovery operation.

Stabiliser codes with dimension $K>0$ have a set of logical operators that allow computation directly on the encoded states. The logical operators are defined as operators that commute with all of the code stabilisers, but are themselves not stabilisers. The encoded quantum information is lost when the code is subject to a Pauli error that is equivalent to a non-trivial logical operator.  The minimum weight of a logical operator therefore gives the distance $D$ of a quantum code. In general, QEC codes are labelled in terms of an $[[N,K,D]]$ notation, where the $N$ is total number of qubits, $K$ is the number of encoded logical qubits and $D$ is the code distance. Note the use of double brackets to differentiate quantum codes from classical codes.

There is a useful mapping for Pauli operators that allows stabiliser QEC codes to be expressed in terms of a binary representation. For single Pauli operators this mapping acts as follows: $\openone\mapsto\left(0|0\right)$, $X\mapsto\left(1|0\right)$, $Y\mapsto\left(1|1\right)$ and $Z\mapsto\left(0|1\right)$. In general, an $n$-qubit Pauli operator $G$ can be mapped to $2N$-bit binary vector of the form
\begin{equation}
	G\mapsto \textbf{g}= \left(\textbf{g}_x|\textbf{g}_z\right)\in\mathbb{F}_2^{2N}\rm,
\end{equation}
where $\textbf{g}_x$ is a $n$-bit binary vector set to $1$ in the qubit locations where $G$ acts with an $X$-operator. Similarly, $\textbf{g}_z$ marks the qubit locations where $G$ acts with a $Z$ operator. A $Y$-Pauli operator acting on qubit $i$ results in a $1$ in the corresponding position of both vectors $\textbf{g}_x$ and $\textbf{g}_z$. As an example, the two-qubit operator $G=X_1Y_2$ maps to the binary representation $\textbf{g}=\left(11|01\right)$.

Using the binary representation, a QEC code can be written in terms of an $M{\times}2N$ parity matrix, $H_Q=\left[H_X|H_Z\right]\in\mathbb{F}_2^{M{\times}2N}$, where each row corresponds to one of the $M$ stabiliser generators. For an error $\textbf{e}_Q=\left(\textbf{e}_X|\textbf{e}_Z\right)$ the code syndrome is given by
\begin{equation}\label{eq:qsynd}
\begin{split}
\textbf{s}_Q=\left[H_X| H_Z\right]\cdot\left(\textbf{e}_Z|\textbf{e}_X\right)=H_X\cdot \textbf{e}_Z+H_Z\cdot \textbf{e}_X\rm.
\end{split}
\end{equation}
In the following sections we will see the binary representation of stabiliser codes facilitates the re-purposing of existing classical codes into quantum codes. Another useful feature of the binary representation is that it enables the straightforward verification of various code properties. For example, the condition that code stabilisers commute amounts to verifying that $H_X$ and $H_Z$ satisfy the following constraint
\begin{equation}\label{eq:commutivity}
    H_X\cdot H_Z^T +H_Z\cdot H_X^T = 0\rm.
\end{equation}

\subsection{Calderbank, Shor and Steane (CSS) codes}

The Calderbank, Shor and Steane (CSS) family is a subset of QEC codes that have disjoint $X$- and $Z$-type stabilisers such that each stabiliser is made up either exclusively of $X$-type Pauli operators, or exclusively of $Z$-type Pauli operators \cite{Calderbank95,Steane96b,Steane97}. In the binary representation, the parity matrix of a CSS code has the form
\begin{equation}\label{eq:h_css}
	H_{\rm CSS}=\left[\begin{array}{@{} c|c @{}}
		0&H'_Z\\
		H'_X&0
	\end{array}\right]\rm.
\end{equation} 
For CSS codes, the stabiliser commutativity constraint defined in Eq.~(\ref{eq:commutivity}) simplifies to verifying that $H'_X\cdot {H'}_Z^T=0$. The structure of CSS codes allows bit and phase-flips to be corrected using separate classical codes. However, the commutativity constraint on stabiliser codes means that it is not simply possible to re-purpose any classical code as the $H'_X$ and $H'_Z$ matrices of a CSS code. Various constructions exist that allow classical codes to be converted to CSS codes. In this work, we focus on two such constructions: the hypergraph product \cite{tillich2013quantum} and the lifted product \cite{Panteleev_2019,panteleev2020quantum}.

\subsection{Hypergraph product codes}

The hypergraph product is a method for creating a quantum code from any two classical \textit{seed} codes \cite{tillich2013quantum,quintavalle_reshape}. To construct a hypergraph product code, we first choose the seed codes: $H_1\in\mathbb{F}_2^{m_1\times n_1}$ with parameters $[n_1,k_1,d_1]$ and $H_2\in\mathbb{F}_2^{m_2\times n_2}$ with parameters $[n_2,k_2,d_2]$. For each seed code $H\in\mathbb{F}_2^{m\times n}$, we also define a transpose code with parity check matrix $H^T\in\mathbb{F}_2^{n\times m}$ and parameters $[m,k^T,d^T]$. Given seed codes $H_1$ and $H_2$, the parity check matrix of the hypergraph product code is as follows
\begin{equation}\label{eq:hgp_css}
\arraycolsep=4pt\def\arraystretch{1}
\thickmuskip=0.5\thickmuskip
\begin{split}
    &H_{HGP}=\left[H_X|H_Z\right]\\
    &=\left[\begin{array}{@{} cc|cc @{}}
    0 & 0 & \openone_{n_1}{\otimes} H_2 & H_1^T{\otimes} \openone_{m_2} \\
    H_1{\otimes}\openone_{n_2} & \openone_{m_1}{\otimes} H_2^T & 0 & 0
    \end{array}
    \right]\rm.
    \end{split}
\end{equation}
The above code $H_{HGP}$ has parameters $[[N,K,D]]$ where
$N=n_1n_2+m_1m_2$, $K=k_1k_2 +k_1^Tk_2^T$ and $D={\rm min}(d_1,d_2,d_1^T,d_2^T)$. The specific advantage of the hypergraph product is that any pair of classical codes can be used as the seed matrices: it is straightforward to verify that the stabiliser commutivity constraint holds for all pairs $H_1$ and $H_2$. Another useful property is that the hypergraph product preserves the sparsity of the seed codes (up to a constant additive factor). The hypergraph product of two LDPC codes therefore yields a quantum code with an LDPC parity check matrix.

Both the $H_X$ and $H_Z$ components of the matrix in Eq.~(\ref{eq:hgp_css}) are formed from a concatenation of two blocks of qubits. We refer to the first block of $n_1n_2$ qubits as \textit{sector one} and the second block of $m_1m_2$ qubits as \textit{sector two}.

\begin{example} \label{example:toric_code}

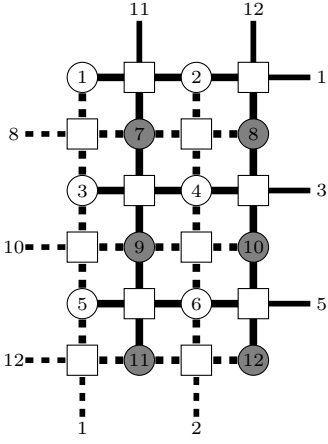
\begin{figure} \label{fig:toric_code}
    \centering
    \begin{tikzpicture}
\draw[black,thick,dashed,line width=1.00mm] (0.0,3.75) -- (0.0,3.0);
\draw[black,thick,dashed,line width=1.00mm] (0.0,2.25) -- (0.0,3.0);
\draw[black,thick,dashed,line width=1.00mm] (0.75,3.0) -- (0.0,3.0);
\draw[black,thick,dashed,line width=0.75mm] (-0.75,3.0) -- (0.0,3.0);
\node () at (-0.900000,3.000000)  [] {\scriptsize 8};
\draw[black,thick,dashed,line width=1.00mm] (1.5,3.75) -- (1.5,3.0);
\draw[black,thick,dashed,line width=1.00mm] (1.5,2.25) -- (1.5,3.0);
\draw[black,thick,dashed,line width=1.00mm] (0.75,3.0) -- (1.5,3.0);
\draw[black,thick,dashed,line width=1.00mm] (2.25,3.0) -- (1.5,3.0);
\draw[black,thick,dashed,line width=1.00mm] (0.0,2.25) -- (0.0,1.5);
\draw[black,thick,dashed,line width=1.00mm] (0.0,0.75) -- (0.0,1.5);
\draw[black,thick,dashed,line width=1.00mm] (0.75,1.5) -- (0.0,1.5);
\draw[black,thick,dashed,line width=0.75mm] (-0.75,1.5) -- (0.0,1.5);
\node () at (-0.900000,1.500000)  [] {\scriptsize 10};
\draw[black,thick,dashed,line width=1.00mm] (1.5,2.25) -- (1.5,1.5);
\draw[black,thick,dashed,line width=1.00mm] (1.5,0.75) -- (1.5,1.5);
\draw[black,thick,dashed,line width=1.00mm] (0.75,1.5) -- (1.5,1.5);
\draw[black,thick,dashed,line width=1.00mm] (2.25,1.5) -- (1.5,1.5);
\draw[black,thick,dashed,line width=0.75mm] (0.0,-0.75) -- (0.0,0.0);
\node () at (0.000000,-0.900000)  [] {\scriptsize 1};
\draw[black,thick,dashed,line width=1.00mm] (0.0,0.75) -- (0.0,0.0);
\draw[black,thick,dashed,line width=1.00mm] (0.75,0.0) -- (0.0,0.0);
\draw[black,thick,dashed,line width=0.75mm] (-0.75,0.0) -- (0.0,0.0);
\node () at (-0.900000,0.000000)  [] {\scriptsize 12};
\draw[black,thick,dashed,line width=0.75mm] (1.5,-0.75) -- (1.5,0.0);
\node () at (1.500000,-0.900000)  [] {\scriptsize 2};
\draw[black,thick,dashed,line width=1.00mm] (1.5,0.75) -- (1.5,0.0);
\draw[black,thick,dashed,line width=1.00mm] (0.75,0.0) -- (1.5,0.0);
\draw[black,thick,dashed,line width=1.00mm] (2.25,0.0) -- (1.5,0.0);
\draw[black,thick,line width=1.00mm] (0.0,3.75) -- (0.75,3.75);
\draw[black,thick,line width=1.00mm] (1.5,3.75) -- (0.75,3.75);
\draw[black,thick,line width=1.00mm] (0.75,3.0) -- (0.75,3.75);
\draw[black,thick,line width=0.75mm] (0.75,4.5) -- (0.75,3.75);
\node () at (0.750000,4.650000)  [] {\scriptsize 11};
\draw[black,thick,line width=0.75mm] (3.0,3.75) -- (2.25,3.75);
\node () at (3.150000,3.750000)  [] {\scriptsize 1};
\draw[black,thick,line width=1.00mm] (1.5,3.75) -- (2.25,3.75);
\draw[black,thick,line width=1.00mm] (2.25,3.0) -- (2.25,3.75);
\draw[black,thick,line width=0.75mm] (2.25,4.5) -- (2.25,3.75);
\node () at (2.250000,4.650000)  [] {\scriptsize 12};
\draw[black,thick,line width=1.00mm] (0.0,2.25) -- (0.75,2.25);
\draw[black,thick,line width=1.00mm] (1.5,2.25) -- (0.75,2.25);
\draw[black,thick,line width=1.00mm] (0.75,3.0) -- (0.75,2.25);
\draw[black,thick,line width=1.00mm] (0.75,1.5) -- (0.75,2.25);
\draw[black,thick,line width=0.75mm] (3.0,2.25) -- (2.25,2.25);
\node () at (3.150000,2.250000)  [] {\scriptsize 3};
\draw[black,thick,line width=1.00mm] (1.5,2.25) -- (2.25,2.25);
\draw[black,thick,line width=1.00mm] (2.25,3.0) -- (2.25,2.25);
\draw[black,thick,line width=1.00mm] (2.25,1.5) -- (2.25,2.25);
\draw[black,thick,line width=1.00mm] (0.0,0.75) -- (0.75,0.75);
\draw[black,thick,line width=1.00mm] (1.5,0.75) -- (0.75,0.75);
\draw[black,thick,line width=1.00mm] (0.75,1.5) -- (0.75,0.75);
\draw[black,thick,line width=1.00mm] (0.75,0.0) -- (0.75,0.75);
\draw[black,thick,line width=0.75mm] (3.0,0.75) -- (2.25,0.75);
\node () at (3.150000,0.750000)  [] {\scriptsize 5};
\draw[black,thick,line width=1.00mm] (1.5,0.75) -- (2.25,0.75);
\draw[black,thick,line width=1.00mm] (2.25,1.5) -- (2.25,0.75);
\draw[black,thick,line width=1.00mm] (2.25,0.0) -- (2.25,0.75);
\filldraw[fill=white, draw=black] (0.000000,3.750000) circle (0.200000) node [label=center:{$\scriptstyle {1} $}]{};
\filldraw[fill=white, draw=black] (1.500000,3.750000) circle (0.200000) node [label=center:{$\scriptstyle {2} $}]{};
\filldraw[fill=white, draw=black] (0.000000,2.250000) circle (0.200000) node [label=center:{$\scriptstyle {3} $}]{};
\filldraw[fill=white, draw=black] (1.500000,2.250000) circle (0.200000) node [label=center:{$\scriptstyle {4} $}]{};
\filldraw[fill=white, draw=black] (0.000000,0.750000) circle (0.200000) node [label=center:{$\scriptstyle {5} $}]{};
\filldraw[fill=white, draw=black] (1.500000,0.750000) circle (0.200000) node [label=center:{$\scriptstyle {6} $}]{};
\filldraw[fill=gray, draw=black] (0.750000,3.000000) circle (0.200000) node [label=center:{$\scriptstyle {7} $}]{};
\filldraw[fill=gray, draw=black] (2.250000,3.000000) circle (0.200000) node [label=center:{$\scriptstyle {8} $}]{};
\filldraw[fill=gray, draw=black] (0.750000,1.500000) circle (0.200000) node [label=center:{$\scriptstyle {9} $}]{};
\filldraw[fill=gray, draw=black] (2.250000,1.500000) circle (0.200000) node [label=center:{$\scriptstyle {10} $}]{};
\filldraw[fill=gray, draw=black] (0.750000,0.000000) circle (0.200000) node [label=center:{$\scriptstyle {11} $}]{};
\filldraw[fill=gray, draw=black] (2.250000,0.000000) circle (0.200000) node [label=center:{$\scriptstyle {12} $}]{};
\filldraw[fill=white, draw=black] (0.550000,3.550000) rectangle (0.950000,3.950000) node [label=above right:{}]{};\filldraw[fill=white, draw=white] (0.750000,3.750000) circle (0.100000) node [label=center:{$\scriptstyle {} $}]{};
\filldraw[fill=white, draw=black] (2.050000,3.550000) rectangle (2.450000,3.950000) node [label=above right:{}]{};\filldraw[fill=white, draw=white] (2.250000,3.750000) circle (0.100000) node [label=center:{$\scriptstyle {} $}]{};
\filldraw[fill=white, draw=black] (0.550000,2.050000) rectangle (0.950000,2.450000) node [label=above right:{}]{};\filldraw[fill=white, draw=white] (0.750000,2.250000) circle (0.100000) node [label=center:{$\scriptstyle {} $}]{};
\filldraw[fill=white, draw=black] (2.050000,2.050000) rectangle (2.450000,2.450000) node [label=above right:{}]{};\filldraw[fill=white, draw=white] (2.250000,2.250000) circle (0.100000) node [label=center:{$\scriptstyle {} $}]{};
\filldraw[fill=white, draw=black] (0.550000,0.550000) rectangle (0.950000,0.950000) node [label=above right:{}]{};\filldraw[fill=white, draw=white] (0.750000,0.750000) circle (0.100000) node [label=center:{$\scriptstyle {} $}]{};
\filldraw[fill=white, draw=black] (2.050000,0.550000) rectangle (2.450000,0.950000) node [label=above right:{}]{};\filldraw[fill=white, draw=white] (2.250000,0.750000) circle (0.100000) node [label=center:{$\scriptstyle {} $}]{};
\filldraw[fill=white, draw=black] (-0.200000,2.800000) rectangle (0.200000,3.200000) node [label=above right:{}]{};\filldraw[fill=white, draw=white] (0.000000,3.000000) circle (0.100000) node [label=center:{$\scriptstyle {} $}]{};
\filldraw[fill=white, draw=black] (1.300000,2.800000) rectangle (1.700000,3.200000) node [label=above right:{}]{};\filldraw[fill=white, draw=white] (1.500000,3.000000) circle (0.100000) node [label=center:{$\scriptstyle {} $}]{};
\filldraw[fill=white, draw=black] (-0.200000,1.300000) rectangle (0.200000,1.700000) node [label=above right:{}]{};\filldraw[fill=white, draw=white] (0.000000,1.500000) circle (0.100000) node [label=center:{$\scriptstyle {} $}]{};
\filldraw[fill=white, draw=black] (1.300000,1.300000) rectangle (1.700000,1.700000) node [label=above right:{}]{};\filldraw[fill=white, draw=white] (1.500000,1.500000) circle (0.100000) node [label=center:{$\scriptstyle {} $}]{};
\filldraw[fill=white, draw=black] (-0.200000,-0.200000) rectangle (0.200000,0.200000) node [label=above right:{}]{};\filldraw[fill=white, draw=white] (0.000000,0.000000) circle (0.100000) node [label=center:{$\scriptstyle {} $}]{};
\filldraw[fill=white, draw=black] (1.300000,-0.200000) rectangle (1.700000,0.200000) node [label=above right:{}]{};\filldraw[fill=white, draw=white] (1.500000,0.000000) circle (0.100000) node [label=center:{$\scriptstyle {} $}]{};
\end{tikzpicture}
    \caption{A $[[12,2,2]]$ CSS toric code obtained from the hypergraph product of a $d=3$ repetition code ($H_1=H_{\rm rep}^3$) with a $d=2$ repetition code ($H_2=H_{\rm rep}^2$). The circular nodes represent data qubits and the square nodes the check qubits. Solid edges denote $Z$-type Pauli checks and dashed edges $X$-checks. The boundary edges are labelled to indicate which qubits they connect to on the opposite side of the lattice. The subset of data qubits filled with white are in qubit sector 1 (corresponding to the first block of $n_1n_2$ qubits in the hypergraph product) whilst the subset filled in grey are the sector two qubits (corresponding to the second block of $m_1m_2$ qubits in the hypergraph product).}
    \label{fig:toric_code}
\end{figure}

The CSS toric code is constructed from the hypergraph product by choosing closed-loop repetition codes as the seed codes $H_1$ and $H_2$. Fig.~\ref{fig:toric_code} shows the factor graph of a toric code constructed from the hypergraph product of a length-three repetition code ($H_1=H^{\rm rep}_3$) and a length-two repetition code ($H_2=H_2^{\rm rep}$). Here, the circular nodes represent the data qubits and the square nodes the check qubits which measure the stabilisers. The solid edges denote $Z$-type Pauli checks and are drawn according to the adjacency matrix $H_Z$, whilst the dashed edges represent $X$-type checks and are drawn according to $H_X$. Boundary edges are labelled to indicate which qubit they connect to on the opposite side of the lattice. We see that each column of the toric code contains a copy of the seed code $H_1$, whilst each row contains a copy of seed code $H_2$. By convention, we define a toric code lattice to have dimension $n_1\times n_2$. The depicted code has parameters $[[12,2,2]]$ defined on a $3\times 2$ lattice. Repetition codes have square parity check matrices which means that $n_{1,2}=m_{1,2}$ for the seed codes $H_1$ and $H_2$. Consequently, toric codes have an equal number of qubits in both sector one and sector two. In Fig~\ref{fig:toric_code}, the sector one qubits are labelled one to six (filled in white) and the sector two qubits six to twelve (filled in grey). Bigger toric codes can be constructed by using larger repetition codes as the hypergraph seeds.

\end{example}

\begin{example}\label{example:hgp1}
The parity check matrix $H$ given by Eq.~(\ref{eq:mkmn}) in Appendix \ref{app:parity_check_matrices} defines a classical code with parameters $[16,4,6]$. Choosing $H_1=H$ and $H_2=H$, we obtain a hypergraph product code with parameters $[[400,16,6]]$. 
\end{example}

\subsection{Lifted product codes}
The hypergraph product construction described by Eq.~(\ref{eq:hgp_css}) is applied to a pair of seeds codes defined by binary parity check matrices. The lifted product applies the same operation, but to a pair of seed protographs, which are matrices with coefficients in the ring of circulants $\mathbb{A}_L$ as defined in Section~\ref{sec:quasi_cyclic}. Before introducing the lifted product, we describe the transpose operation for elements in $\mathbb{A}_L$. For an element $\alpha=\sum_{i=0}^{L-1}a_i \lambda^i_L$, we define its transpose $\alpha^T$ as
\begin{equation}
\alpha^T\coloneqq\sum_{i=0}^{L-1} a_i \lambda^{(-i)}_L = \sum_{i=0}^{L-1} a_i \lambda^{(L-i)}_L\rm.\end{equation}
For example, if $\alpha=\lambda^0_L+\lambda^1_L$ then $\alpha^T=\lambda^0_L+\lambda^{(-1)}_L$. Given an $m{\times}n$ protograph $A \in \mathbb{A}_L^{m{\times}n}$ with coefficients $A_{ij}\in \mathbb{A}_L$, the transpose is an $n{\times}m$ protograph $A^T \in \mathbb{A}_L^{n{\times}m}$ with coefficients $(A^T)_{ij} = (A_{ji})^T$. \jr{We also define an identity protograph, $E_m\in \mathbb{A}_L^{m\times m}$, as an $m\times m$ diagonal matrix where all non-zero entries are $\lambda_L^0 \in \mathbb{A}_L$ such that
\begin{equation}\label{eq:protograph_identity}
    E_m={\rm diag}(\lambda_L^0,\dotsc,\lambda_L^0)\in \mathbb{A}_L^{m\times m}\rm.
\end{equation}
Mapped to binary, the identity protograph is an size-$m^2$ identity matrix: $\mathcal{B}(E_m)=\openone_{m^2}$. For example, the identity protograph $E_3\in \mathbb{A}_L^{3\times 3}$ is given by
\begin{equation}
    E_3=\begin{bmatrix}
    \lambda_3^0&0&0\\
    0&\lambda_3^0&0\\
    0&0&\lambda_3^0
    \end{bmatrix}\rm,
\end{equation}
which maps to a size-$9$ identity matrix $\openone_9$. The identity protograph $E_m\in \mathbb{A}_L^{m\times m}$ leaves all protographs $A\in \mathbb{A}_L^{m\times n}$ unchanged under matrix multiplication $E_m\cdot A = A$. Finally, we note that the tensor product between two protographs is well defined as a simple extension of the standard tensor product to the ring of circulants: the tensor product between $A\in \mathbb{A}_L^{m\times n}$ and $B\in \mathbb{A}_L^{l\times k}$ yields the protograph $C=A\otimes B \in \mathbb{A}_L^{ml\times nk}$. As an example, the tensor product between $E_2\in \mathbb{A}_L^{2\times 2}$ and $A\in \mathbb{A}_L^{m\times n}$ gives the following block protograph\begin{equation}
 E_2\otimes A =\begin{bmatrix}
 A&0\\0&A 
 \end{bmatrix}  \in \mathbb{A}_L^{2m\times 2n} \rm.
\end{equation}
}

To construct a lifted product code, we first choose two protographs: $A_1 \in \mathbb{A}_L^{m'_1{\times}n'_1}$ with dimensions $m'_1{\times}n'_1$ and $A_2 \in \mathbb{A}_L^{m'_2{\times}n'_2}$ with dimensions $m'_2{\times}n'_2$. Given these protographs, the lifted product defines a quantum protograph $A_Q$ 
\begin{equation}\label{eq:lifted_hgp}
\arraycolsep=4pt\def\arraystretch{1}
\thickmuskip=0.5\thickmuskip
\begin{split}
    &A_{Q}=\left[A_X|A_Z\right]\\
    &=\left[\begin{array}{@{} cc|cc @{}}
    0 & 0 & E_{n'_1}{\otimes} A_2 & (A_1)^T{\otimes} E_{m'_2} \\
    A_1{\otimes}E_{n'_2} & E_{m'_1}{\otimes} (A_2)^T & 0 & 0
    \end{array}
    \right]\rm.
    \end{split}
\end{equation}
Note that this has the same form as the hypergraph product given by Eq.~(\ref{eq:hgp_css}). The CSS parity check matrix corresponding to $A_Q$ is given by
\begin{equation}
\mathfrak{B}(A_Q)\rm,
\end{equation}
where $\mathfrak{B}(.)$ is the mapping defined in Eq.~(\ref{eq:binary_mapping}) that returns the binary representation of a protograph.  The length of the lifted product code is given by $N=L(m'_1m'_2 + n'_1n'_2)$.

\etc{Despite similarities to the hypergraph product, the lifted product is different to the hypergraph product of $\mathfrak{B}(A_1)$ and  $\mathfrak{B}(A_2)$, which would have length $N=L^2(m'_1m'_2 + n'_1n'_2)$.  The crucial difference is that the lifted product applies $\mathfrak{B}(.)$ after forming block matrix of Eq.~(\ref{eq:lifted_hgp}) rather than before.}

\begin{example}\label{example:lpc}
Consider the following protograph
\begin{equation}\label{eq:proto1}
A_L=\begin{bmatrix}
\lambda_L^{0}&\lambda_L^{11}&\lambda_L^{7}&\lambda_L^{12}\\ 
\lambda_L^{1}&\lambda_L^{8}&\lambda_L^{1}&\lambda_L^{8}\\ 
\lambda_L^{11}&\lambda_L^{0}&\lambda_L^{4}&\lambda_L^{8}\\ 
\lambda_L^{6}&\lambda_L^{2}&\lambda_L^{4}&\lambda_L^{12}\\
\end{bmatrix}\rm.
\end{equation}
When $L=13$, $\mathfrak{B}(A_{13})$ is a $(4,4)$-LDPC matrix which yields a code with parameters $[52,3,26]$. We generated this protograph using the progressive edge growth (PEG) technique for quasi-cyclic codes \cite{peg}. The PEG method proceeds by randomly generating protograph elements under the constraint that each new term does not decrease the code girth below a certain target value. Here, the code girth is defined as the minimum-length loop in the code factor graph. \joschka{Factor graphs with high-girth are beneficial as small-length loops can cause decoders based on belief propagation to fail. More specific details concerning decoding methods are discussed in Section~\ref{sec:numerics}.} The parity check matrix $\mathfrak{B}(A_{13})$ has minimum-girth six. The code distance $d=26$ was determined via an exhaustive search over all codewords.

By applying the lifted product to two copies of $A_{13}$, we obtain a quantum code with parameters $[[416,18,d\leq 20]]$. There is no known method of computing the code distance of a \etc{generic} lifted product code given the distances of the seed codes. However, we can obtain a heuristic indication of the code performance by running simulations. The upper-bound $d\leq 20$ on the code distance is an estimate based on the minimum-weight logical operator observed after extensive decoding simulation with the BP+OSD decoder. \joschka{By comparison, the $[[400,16,6]]$ hypergraph product code from Example \ref{example:hgp1} has a similar code rate but a minimum distance of six.}   

\end{example}

\section{Bias-tailored QEC}\label{sec:bias_tailored_qec}

\subsection{Error model}

We now explore how QEC codes can be designed to exploit noise asymmetry. To characterise the qubit bias, we consider an error model in which each qubit is independently subject to a Pauli channel $\mathcal{E}_Q(\rho)$ of the form
\begin{equation}\label{eq:error_channel}
\mathcal{E}_Q(\rho)= (1-p)\rho+p_XX\rho X+p_YY\rho Y+p_ZZ\rho Z\rm,
\end{equation} 
where $\rho$ is the density matrix of the single-qubit state and $p_X$, $p_Y$ and $p_Z$ are the physical error rates for $X\text{-,}$ $Y$- and $Z$-type errors respectively. The total error rate $p$ is given by the sum over the three error types $p=p_X+p_Y+p_Z$. Under this channel, the noise bias $\eta_i$ is defined
\begin{equation}\label{eq:bias}
\eta_i=\frac{p_i}{\sum_{j\neq i} p_j}\rm,
\end{equation}
where $p_i,p_j\in\{p_X,p_Y,p_Z\}$. For example, the $Z$-error bias is $\eta_Z=0.5$ for depolarising noise where $p_X=p_Y=p_Z$. Now consider the case of infinite $Z$-bias $\eta_Z=\infty$ where the $Z$-error probability has some non-zero value $0<p_Z\leq 1$ whilst the $X$- and $Y$-error probabilities are zero $p_X=p_Y=0$. In this regime, the quantum syndrome equation defined in Eq.(\ref{eq:qsynd}) simplifies to $\textbf{s}_Q=H_X\cdot \textbf{e}_Z$, as the probability of obtaining either an $X$- or $Y$-error is zero. The performance of the code therefore depends exclusively on the $H_X$ component of the quantum parity check matrix. Similarly, for infinite $X$-bias $\eta_X=\infty$, the syndrome equation simplifies to $\textbf{s}_Q=H_Z\cdot \textbf{e}_X$ leaving the code performance dependent on $H_Z$. From this, we see that tailoring a QEC code to a specific bias-type involves ensuring that the relevant sub-component of the quantum parity check matrix ($H_X$ or $H_Z$) is itself a \textit{good} code, both in terms of its distance and the ease of its decoding.

\jr{How do we expect the capacity of the quantum error channel to change with increasing bias? A partial answer to this question can be obtained by applying the Shannon noisy coding theorem to obtain a quantity known as the \textit{Hashing threshold}. The Hashing threshold gives a lower bound\footnote{\jr{The Hashing threshold is a lower bound as it does not account for quantum effects such as superposition and entanglement that increase the code threshold.}} on the physical error rate below which a quantum code can in theory arbitrarily suppress the error rate. For the quantum channel defined in Eq.~(\ref{eq:error_channel}), the Hashing threshold for depolarising noise is $18.9\%$. As the bias is increased (for $X$-,$Y$- or $Z$-noise), the Hashing threshold increases rapidly until converging to a value of $50\%$ in the limit of infinite-bias. The Hashing threshold provides the theoretical motivation for bias-tailoring: the higher the bias, the higher the channel capacity. A more detailed discussion of the Hashing threshold can be found in Appendix~\ref{app:hashing}.  
}

\subsection{The CSS toric code under biased noise}

\begin{figure}
    \centering
    \begin{tikzpicture}
\draw[black,thick,line width=1.00mm] (0.0,3.75) -- (0.75,3.75);
\draw[black,thick,line width=1.00mm] (1.5,3.75) -- (0.75,3.75);
\draw[black,thick,line width=1.00mm] (0.75,3.0) -- (0.75,3.75);
\draw[black,thick,line width=0.75mm] (0.75,4.5) -- (0.75,3.75);
\node () at (0.750000,4.650000)  [] {\scriptsize 11};
\draw[black,thick,line width=0.75mm] (3.0,3.75) -- (2.25,3.75);
\node () at (3.150000,3.750000)  [] {\scriptsize 1};
\draw[black,thick,line width=1.00mm] (1.5,3.75) -- (2.25,3.75);
\draw[black,thick,line width=1.00mm] (2.25,3.0) -- (2.25,3.75);
\draw[black,thick,line width=0.75mm] (2.25,4.5) -- (2.25,3.75);
\node () at (2.250000,4.650000)  [] {\scriptsize 12};
\draw[black,thick,line width=1.00mm] (0.0,2.25) -- (0.75,2.25);
\draw[black,thick,line width=1.00mm] (1.5,2.25) -- (0.75,2.25);
\draw[black,thick,line width=1.00mm] (0.75,3.0) -- (0.75,2.25);
\draw[black,thick,line width=1.00mm] (0.75,1.5) -- (0.75,2.25);
\draw[black,thick,line width=0.75mm] (3.0,2.25) -- (2.25,2.25);
\node () at (3.150000,2.250000)  [] {\scriptsize 3};
\draw[black,thick,line width=1.00mm] (1.5,2.25) -- (2.25,2.25);
\draw[black,thick,line width=1.00mm] (2.25,3.0) -- (2.25,2.25);
\draw[black,thick,line width=1.00mm] (2.25,1.5) -- (2.25,2.25);
\draw[black,thick,line width=1.00mm] (0.0,0.75) -- (0.75,0.75);
\draw[black,thick,line width=1.00mm] (1.5,0.75) -- (0.75,0.75);
\draw[black,thick,line width=1.00mm] (0.75,1.5) -- (0.75,0.75);
\draw[black,thick,line width=1.00mm] (0.75,0.0) -- (0.75,0.75);
\draw[black,thick,line width=0.75mm] (3.0,0.75) -- (2.25,0.75);
\node () at (3.150000,0.750000)  [] {\scriptsize 5};
\draw[black,thick,line width=1.00mm] (1.5,0.75) -- (2.25,0.75);
\draw[black,thick,line width=1.00mm] (2.25,1.5) -- (2.25,0.75);
\draw[black,thick,line width=1.00mm] (2.25,0.0) -- (2.25,0.75);
\filldraw[fill=white, draw=black] (0.000000,3.750000) circle (0.200000) node [label=center:{$\scriptstyle {1} $}]{};
\filldraw[fill=white, draw=black] (1.500000,3.750000) circle (0.200000) node [label=center:{$\scriptstyle {2} $}]{};
\filldraw[fill=red, draw=black] (0.000000,2.250000) circle (0.200000) node [label=center:{$\scriptstyle {3} $}]{};
\filldraw[fill=red, draw=black] (1.500000,2.250000) circle (0.200000) node [label=center:{$\scriptstyle {4} $}]{};
\filldraw[fill=white, draw=black] (0.000000,0.750000) circle (0.200000) node [label=center:{$\scriptstyle {5} $}]{};
\filldraw[fill=white, draw=black] (1.500000,0.750000) circle (0.200000) node [label=center:{$\scriptstyle {6} $}]{};
\filldraw[fill=white, draw=black] (0.750000,3.000000) circle (0.200000) node [label=center:{$\scriptstyle {7} $}]{};
\filldraw[fill=white, draw=black] (2.250000,3.000000) circle (0.200000) node [label=center:{$\scriptstyle {8} $}]{};
\filldraw[fill=white, draw=black] (0.750000,1.500000) circle (0.200000) node [label=center:{$\scriptstyle {9} $}]{};
\filldraw[fill=white, draw=black] (2.250000,1.500000) circle (0.200000) node [label=center:{$\scriptstyle {10} $}]{};
\filldraw[fill=white, draw=black] (0.750000,0.000000) circle (0.200000) node [label=center:{$\scriptstyle {11} $}]{};
\filldraw[fill=white, draw=black] (2.250000,0.000000) circle (0.200000) node [label=center:{$\scriptstyle {12} $}]{};
\filldraw[fill=white, draw=black] (0.550000,3.550000) rectangle (0.950000,3.950000) node [label=above right:{}]{};\filldraw[fill=white, draw=white] (0.750000,3.750000) circle (0.100000) node [label=center:{$\scriptstyle {} $}]{};
\filldraw[fill=white, draw=black] (2.050000,3.550000) rectangle (2.450000,3.950000) node [label=above right:{}]{};\filldraw[fill=white, draw=white] (2.250000,3.750000) circle (0.100000) node [label=center:{$\scriptstyle {} $}]{};
\filldraw[fill=white, draw=black] (0.550000,2.050000) rectangle (0.950000,2.450000) node [label=above right:{}]{};\filldraw[fill=white, draw=white] (0.750000,2.250000) circle (0.100000) node [label=center:{$\scriptstyle {} $}]{};
\filldraw[fill=white, draw=black] (2.050000,2.050000) rectangle (2.450000,2.450000) node [label=above right:{}]{};\filldraw[fill=white, draw=white] (2.250000,2.250000) circle (0.100000) node [label=center:{$\scriptstyle {} $}]{};
\filldraw[fill=white, draw=black] (0.550000,0.550000) rectangle (0.950000,0.950000) node [label=above right:{}]{};\filldraw[fill=white, draw=white] (0.750000,0.750000) circle (0.100000) node [label=center:{$\scriptstyle {} $}]{};
\filldraw[fill=white, draw=black] (2.050000,0.550000) rectangle (2.450000,0.950000) node [label=above right:{}]{};\filldraw[fill=white, draw=white] (2.250000,0.750000) circle (0.100000) node [label=center:{$\scriptstyle {} $}]{};
\filldraw[fill=white, draw=black] (-0.200000,2.800000) rectangle (0.200000,3.200000) node [label=above right:{}]{};\filldraw[fill=white, draw=white] (0.000000,3.000000) circle (0.100000) node [label=center:{$\scriptstyle {} $}]{};
\filldraw[fill=white, draw=black] (1.300000,2.800000) rectangle (1.700000,3.200000) node [label=above right:{}]{};\filldraw[fill=white, draw=white] (1.500000,3.000000) circle (0.100000) node [label=center:{$\scriptstyle {} $}]{};
\filldraw[fill=white, draw=black] (-0.200000,1.300000) rectangle (0.200000,1.700000) node [label=above right:{}]{};\filldraw[fill=white, draw=white] (0.000000,1.500000) circle (0.100000) node [label=center:{$\scriptstyle {} $}]{};
\filldraw[fill=white, draw=black] (1.300000,1.300000) rectangle (1.700000,1.700000) node [label=above right:{}]{};\filldraw[fill=white, draw=white] (1.500000,1.500000) circle (0.100000) node [label=center:{$\scriptstyle {} $}]{};
\filldraw[fill=white, draw=black] (-0.200000,-0.200000) rectangle (0.200000,0.200000) node [label=above right:{}]{};\filldraw[fill=white, draw=white] (0.000000,0.000000) circle (0.100000) node [label=center:{$\scriptstyle {} $}]{};
\filldraw[fill=white, draw=black] (1.300000,-0.200000) rectangle (1.700000,0.200000) node [label=above right:{}]{};\filldraw[fill=white, draw=white] (1.500000,0.000000) circle (0.100000) node [label=center:{$\scriptstyle {} $}]{};
\end{tikzpicture}
    \caption{The $H_Z$ component of the $[[12,2,2]]$ CSS toric code depicted in Fig.~\ref{fig:toric_code}. The nodes filled with red show a weight-two logical operator that spans the lattice.}
    \label{fig:toric_hz}
\end{figure}
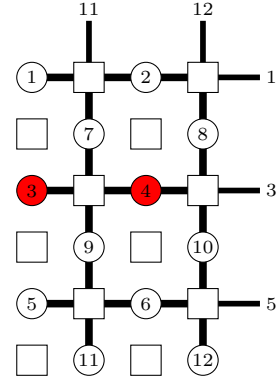

We first consider the behaviour of the standard CSS code (as presented in Example \ref{example:toric_code}) in different bias regimes. Under depolarising noise the upper bound on the threshold of the CSS toric code is $18.9\%$ as determined by mapping the code Hamiltonian to the classically disordered eight-bond Ising model \cite{bombin2012strong}. Thresholds close to this upper bound have been numerically observed via Monte Carlo simulations of the toric code using decoders based on minimum-weight perfect matching \cite{Dennis02,Criger18}. From this we see that under depolarising noise, the toric code matches the Hashing threshold of $18.9\%$. \jr{However, as the bias is increased, the toric code's threshold begins to drop counter to what we would expect from the Hashing threshold. To explore why this occurs,} we consider the infinite $X$-bias regime where the performance of the toric code relies exclusively on the measurement and decoding of its $Z$-stabilisers. Fig.~\ref{fig:toric_hz} shows the factor graph of the $H_Z$ subcomponent of the $[[12,2,2]]$ toric code introduced in Example \ref{example:toric_code}. The nodes shaded in red depict a weight-two $X$-logical operator that spans the lattice. In the limit of infinite $X$-bias, this toric code therefore has distance $d_X=2$. By again mapping to a statistical physics model, the threshold of $H_Z$ (and $H_X$) component of the toric code under infinite-bias is upper bounded at $10.9\%$ \cite{bombin2012strong}. This is considerably lower than the Hashing threshold of $50\%$ in the infinite-bias regime. It is here that we see the weakness of the CSS toric code when it comes asymmetric channels: its $H_Z$ and $H_X$ sub-components are themselves not good classical codes. Consequently, the CSS toric code is unable to leverage the increase in channel capacity that comes with increased bias. 

\subsection{The XZZX toric code}

The XZZX surface code is a modified version of the CSS surface code  that has recently been shown to perform exceptionally well under biased-noise \cite{wen2002,kovalev12,ataides2020}. The XZZX surface code, as presented in \cite{wen2002,kovalev12,ataides2020}, is defined on a rotated lattice with periodic boundary conditions. The improved performance of the XZZX surface code relative to the standard CSS code can be attributed to two factors: 1) a redefinition of the stabilisers that results in $H_X$ and $H_Z$ sub-components reducing to a set of decoupled repetition codes with a threshold of $50\%$; 2) \textit{twisted} periodic boundary conditions that lead to longer logical operators in the infinite-bias limit. \joschkaedit{In this paper, we explore an XZZX protocol based on the CSS toric code and defined on a rectangular lattice.} In this setting, we will see that both XZZX modifications -- stabiliser redefinition and boundary twists -- can be directly derived from the lifted product.

\begin{figure}
    \centering
    \begin{tikzpicture}
\draw[black,thick,dashed,line width=1.00mm] (0.75,3.0) -- (0.75,3.75);
\draw[black,thick,dashed,line width=0.75mm] (0.75,4.5) -- (0.75,3.75);
\node () at (0.750000,4.650000)  [] {\scriptsize 11};
\draw[black,thick,dashed,line width=1.00mm] (2.25,3.0) -- (2.25,3.75);
\draw[black,thick,dashed,line width=0.75mm] (2.25,4.5) -- (2.25,3.75);
\node () at (2.250000,4.650000)  [] {\scriptsize 12};
\draw[black,thick,dashed,line width=1.00mm] (0.75,3.0) -- (0.75,2.25);
\draw[black,thick,dashed,line width=1.00mm] (0.75,1.5) -- (0.75,2.25);
\draw[black,thick,dashed,line width=1.00mm] (2.25,3.0) -- (2.25,2.25);
\draw[black,thick,dashed,line width=1.00mm] (2.25,1.5) -- (2.25,2.25);
\draw[black,thick,dashed,line width=1.00mm] (0.75,1.5) -- (0.75,0.75);
\draw[black,thick,dashed,line width=1.00mm] (0.75,0.0) -- (0.75,0.75);
\draw[black,thick,dashed,line width=1.00mm] (2.25,1.5) -- (2.25,0.75);
\draw[black,thick,dashed,line width=1.00mm] (2.25,0.0) -- (2.25,0.75);
\draw[black,thick,dashed,line width=1.00mm] (0.0,3.75) -- (0.0,3.0);
\draw[black,thick,dashed,line width=1.00mm] (0.0,2.25) -- (0.0,3.0);
\draw[black,thick,dashed,line width=1.00mm] (1.5,3.75) -- (1.5,3.0);
\draw[black,thick,dashed,line width=1.00mm] (1.5,2.25) -- (1.5,3.0);
\draw[black,thick,dashed,line width=1.00mm] (0.0,2.25) -- (0.0,1.5);
\draw[black,thick,dashed,line width=1.00mm] (0.0,0.75) -- (0.0,1.5);
\draw[black,thick,dashed,line width=1.00mm] (1.5,2.25) -- (1.5,1.5);
\draw[black,thick,dashed,line width=1.00mm] (1.5,0.75) -- (1.5,1.5);
\draw[black,thick,dashed,line width=0.75mm] (0.0,-0.75) -- (0.0,0.0);
\node () at (0.000000,-0.900000)  [] {\scriptsize 1};
\draw[black,thick,dashed,line width=1.00mm] (0.0,0.75) -- (0.0,0.0);
\draw[black,thick,dashed,line width=0.75mm] (1.5,-0.75) -- (1.5,0.0);
\node () at (1.500000,-0.900000)  [] {\scriptsize 2};
\draw[black,thick,dashed,line width=1.00mm] (1.5,0.75) -- (1.5,0.0);
\draw[black,thick,line width=1.00mm] (0.0,3.75) -- (0.75,3.75);
\draw[black,thick,line width=1.00mm] (1.5,3.75) -- (0.75,3.75);
\draw[black,thick,line width=0.75mm] (3.0,3.75) -- (2.25,3.75);
\node () at (3.150000,3.750000)  [] {\scriptsize 1};
\draw[black,thick,line width=1.00mm] (1.5,3.75) -- (2.25,3.75);
\draw[black,thick,line width=1.00mm] (0.0,2.25) -- (0.75,2.25);
\draw[black,thick,line width=1.00mm] (1.5,2.25) -- (0.75,2.25);
\draw[black,thick,line width=0.75mm] (3.0,2.25) -- (2.25,2.25);
\node () at (3.150000,2.250000)  [] {\scriptsize 3};
\draw[black,thick,line width=1.00mm] (1.5,2.25) -- (2.25,2.25);
\draw[black,thick,line width=1.00mm] (0.0,0.75) -- (0.75,0.75);
\draw[black,thick,line width=1.00mm] (1.5,0.75) -- (0.75,0.75);
\draw[black,thick,line width=0.75mm] (3.0,0.75) -- (2.25,0.75);
\node () at (3.150000,0.750000)  [] {\scriptsize 5};
\draw[black,thick,line width=1.00mm] (1.5,0.75) -- (2.25,0.75);
\draw[black,thick,line width=1.00mm] (0.75,3.0) -- (0.0,3.0);
\draw[black,thick,line width=0.75mm] (-0.75,3.0) -- (0.0,3.0);
\node () at (-0.900000,3.000000)  [] {\scriptsize 8};
\draw[black,thick,line width=1.00mm] (0.75,3.0) -- (1.5,3.0);
\draw[black,thick,line width=1.00mm] (2.25,3.0) -- (1.5,3.0);
\draw[black,thick,line width=1.00mm] (0.75,1.5) -- (0.0,1.5);
\draw[black,thick,line width=0.75mm] (-0.75,1.5) -- (0.0,1.5);
\node () at (-0.900000,1.500000)  [] {\scriptsize 10};
\draw[black,thick,line width=1.00mm] (0.75,1.5) -- (1.5,1.5);
\draw[black,thick,line width=1.00mm] (2.25,1.5) -- (1.5,1.5);
\draw[black,thick,line width=1.00mm] (0.75,0.0) -- (0.0,0.0);
\draw[black,thick,line width=0.75mm] (-0.75,0.0) -- (0.0,0.0);
\node () at (-0.900000,0.000000)  [] {\scriptsize 12};
\draw[black,thick,line width=1.00mm] (0.75,0.0) -- (1.5,0.0);
\draw[black,thick,line width=1.00mm] (2.25,0.0) -- (1.5,0.0);
\filldraw[fill=white, draw=black] (0.000000,3.750000) circle (0.200000) node [label=center:{$\scriptstyle {1} $}]{};
\filldraw[fill=white, draw=black] (1.500000,3.750000) circle (0.200000) node [label=center:{$\scriptstyle {2} $}]{};
\filldraw[fill=white, draw=black] (0.000000,2.250000) circle (0.200000) node [label=center:{$\scriptstyle {3} $}]{};
\filldraw[fill=white, draw=black] (1.500000,2.250000) circle (0.200000) node [label=center:{$\scriptstyle {4} $}]{};
\filldraw[fill=white, draw=black] (0.000000,0.750000) circle (0.200000) node [label=center:{$\scriptstyle {5} $}]{};
\filldraw[fill=white, draw=black] (1.500000,0.750000) circle (0.200000) node [label=center:{$\scriptstyle {6} $}]{};
\filldraw[fill=white, draw=black] (0.750000,3.000000) circle (0.200000) node [label=center:{$\scriptstyle {7} $}]{};
\filldraw[fill=white, draw=black] (2.250000,3.000000) circle (0.200000) node [label=center:{$\scriptstyle {8} $}]{};
\filldraw[fill=white, draw=black] (0.750000,1.500000) circle (0.200000) node [label=center:{$\scriptstyle {9} $}]{};
\filldraw[fill=white, draw=black] (2.250000,1.500000) circle (0.200000) node [label=center:{$\scriptstyle {10} $}]{};
\filldraw[fill=white, draw=black] (0.750000,0.000000) circle (0.200000) node [label=center:{$\scriptstyle {11} $}]{};
\filldraw[fill=white, draw=black] (2.250000,0.000000) circle (0.200000) node [label=center:{$\scriptstyle {12} $}]{};
\filldraw[fill=white, draw=black] (0.550000,3.550000) rectangle (0.950000,3.950000) node [label=above right:{}]{};\filldraw[fill=white, draw=white] (0.750000,3.750000) circle (0.100000) node [label=center:{$\scriptstyle {} $}]{};
\filldraw[fill=white, draw=black] (2.050000,3.550000) rectangle (2.450000,3.950000) node [label=above right:{}]{};\filldraw[fill=white, draw=white] (2.250000,3.750000) circle (0.100000) node [label=center:{$\scriptstyle {} $}]{};
\filldraw[fill=white, draw=black] (0.550000,2.050000) rectangle (0.950000,2.450000) node [label=above right:{}]{};\filldraw[fill=white, draw=white] (0.750000,2.250000) circle (0.100000) node [label=center:{$\scriptstyle {} $}]{};
\filldraw[fill=white, draw=black] (2.050000,2.050000) rectangle (2.450000,2.450000) node [label=above right:{}]{};\filldraw[fill=white, draw=white] (2.250000,2.250000) circle (0.100000) node [label=center:{$\scriptstyle {} $}]{};
\filldraw[fill=white, draw=black] (0.550000,0.550000) rectangle (0.950000,0.950000) node [label=above right:{}]{};\filldraw[fill=white, draw=white] (0.750000,0.750000) circle (0.100000) node [label=center:{$\scriptstyle {} $}]{};
\filldraw[fill=white, draw=black] (2.050000,0.550000) rectangle (2.450000,0.950000) node [label=above right:{}]{};\filldraw[fill=white, draw=white] (2.250000,0.750000) circle (0.100000) node [label=center:{$\scriptstyle {} $}]{};
\filldraw[fill=white, draw=black] (-0.200000,2.800000) rectangle (0.200000,3.200000) node [label=above right:{}]{};\filldraw[fill=white, draw=white] (0.000000,3.000000) circle (0.100000) node [label=center:{$\scriptstyle {} $}]{};
\filldraw[fill=white, draw=black] (1.300000,2.800000) rectangle (1.700000,3.200000) node [label=above right:{}]{};\filldraw[fill=white, draw=white] (1.500000,3.000000) circle (0.100000) node [label=center:{$\scriptstyle {} $}]{};
\filldraw[fill=white, draw=black] (-0.200000,1.300000) rectangle (0.200000,1.700000) node [label=above right:{}]{};\filldraw[fill=white, draw=white] (0.000000,1.500000) circle (0.100000) node [label=center:{$\scriptstyle {} $}]{};
\filldraw[fill=white, draw=black] (1.300000,1.300000) rectangle (1.700000,1.700000) node [label=above right:{}]{};\filldraw[fill=white, draw=white] (1.500000,1.500000) circle (0.100000) node [label=center:{$\scriptstyle {} $}]{};
\filldraw[fill=white, draw=black] (-0.200000,-0.200000) rectangle (0.200000,0.200000) node [label=above right:{}]{};\filldraw[fill=white, draw=white] (0.000000,0.000000) circle (0.100000) node [label=center:{$\scriptstyle {} $}]{};
\filldraw[fill=white, draw=black] (1.300000,-0.200000) rectangle (1.700000,0.200000) node [label=above right:{}]{};\filldraw[fill=white, draw=white] (1.500000,0.000000) circle (0.100000) node [label=center:{$\scriptstyle {} $}]{};
\end{tikzpicture}
    \caption{The $[[12,2,2]]$ XZZX toric code obtained after Hadamard rotating the CSS toric code. This code is non-CSS where each check measures the same XZZX stabiliser on its adjacent qubits.}
    \label{fig:xzzx_toric_notwist}
\end{figure}
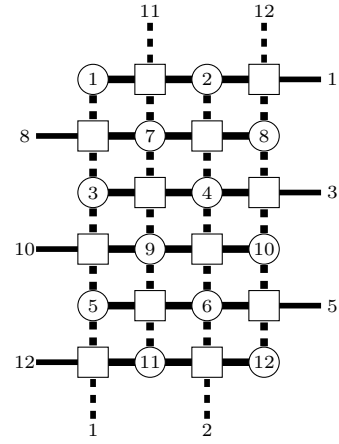

To construct a XZZX code, the first step is to redefine the stabilisers of the hypergraph product via an operation we refer to as a \textit{Hadmard rotation}. The Hadmard rotation applies a Hadamard gate to all sector two qubits. The effect of this is that all $X$-check edges incident on sector two qubits are transformed to $Z$-check edges and vice-versa. Following the Hadmard rotation, the hypergraph product defined in Eq.~(\ref{eq:hgp_css}) has the form
\begin{equation}\label{eq:hgp_xzzx}
\arraycolsep=4pt\def\arraystretch{1}
\thickmuskip=0.5\thickmuskip
\begin{split}
    &H'_{HGP}=\left[H_X|H_Z\right]\\
    &=\left[\begin{array}{@{} cc|cc @{}}
    0 & H_1^T{\otimes} \openone_{m_2} & \openone_{n_1}{\otimes} H_2 & 0 \\
    H_1{\otimes}\openone_{n_2} & 0 & 0 & \openone_{m_1}{\otimes} H_2^T
    \end{array}
    \right]\rm.
    \end{split}
\end{equation}
The above parity check matrix has mixed-type stabilisers, and is therefore no longer a CSS code. Fig.~\ref{fig:xzzx_toric_notwist} shows the factor graph of the XZZX toric code obtained by applying a Hadamard rotation to the $[[12,2,2]]$ CSS toric code depicted in Fig.~\ref{fig:toric_code}. The key difference is that all $Z$-check edges now act horizontally whilst all $X$-check edges act vertically. Every parity check qubit measures the same XZZX stabiliser which lends the code its name.

\begin{figure}
    \centering
    \begin{tikzpicture}
\draw[black,thick,line width=1.00mm] (0.0,3.75) -- (0.75,3.75);
\draw[black,thick,line width=1.00mm] (1.5,3.75) -- (0.75,3.75);
\draw[black,thick,line width=0.75mm] (3.0,3.75) -- (2.25,3.75);
\node () at (3.150000,3.750000)  [] {\scriptsize 1};
\draw[black,thick,line width=1.00mm] (1.5,3.75) -- (2.25,3.75);
\draw[black,thick,line width=1.00mm] (0.0,2.25) -- (0.75,2.25);
\draw[black,thick,line width=1.00mm] (1.5,2.25) -- (0.75,2.25);
\draw[black,thick,line width=0.75mm] (3.0,2.25) -- (2.25,2.25);
\node () at (3.150000,2.250000)  [] {\scriptsize 3};
\draw[black,thick,line width=1.00mm] (1.5,2.25) -- (2.25,2.25);
\draw[black,thick,line width=1.00mm] (0.0,0.75) -- (0.75,0.75);
\draw[black,thick,line width=1.00mm] (1.5,0.75) -- (0.75,0.75);
\draw[black,thick,line width=0.75mm] (3.0,0.75) -- (2.25,0.75);
\node () at (3.150000,0.750000)  [] {\scriptsize 5};
\draw[black,thick,line width=1.00mm] (1.5,0.75) -- (2.25,0.75);
\draw[black,thick,line width=1.00mm] (0.75,3.0) -- (0.0,3.0);
\draw[black,thick,line width=0.75mm] (-0.75,3.0) -- (0.0,3.0);
\node () at (-0.900000,3.000000)  [] {\scriptsize 8};
\draw[black,thick,line width=1.00mm] (0.75,3.0) -- (1.5,3.0);
\draw[black,thick,line width=1.00mm] (2.25,3.0) -- (1.5,3.0);
\draw[black,thick,line width=1.00mm] (0.75,1.5) -- (0.0,1.5);
\draw[black,thick,line width=0.75mm] (-0.75,1.5) -- (0.0,1.5);
\node () at (-0.900000,1.500000)  [] {\scriptsize 10};
\draw[black,thick,line width=1.00mm] (0.75,1.5) -- (1.5,1.5);
\draw[black,thick,line width=1.00mm] (2.25,1.5) -- (1.5,1.5);
\draw[black,thick,line width=1.00mm] (0.75,0.0) -- (0.0,0.0);
\draw[black,thick,line width=0.75mm] (-0.75,0.0) -- (0.0,0.0);
\node () at (-0.900000,0.000000)  [] {\scriptsize 12};
\draw[black,thick,line width=1.00mm] (0.75,0.0) -- (1.5,0.0);
\draw[black,thick,line width=1.00mm] (2.25,0.0) -- (1.5,0.0);
\filldraw[fill=white, draw=black] (0.000000,3.750000) circle (0.200000) node [label=center:{$\scriptstyle {1} $}]{};
\filldraw[fill=white, draw=black] (1.500000,3.750000) circle (0.200000) node [label=center:{$\scriptstyle {2} $}]{};
\filldraw[fill=red, draw=black] (0.000000,2.250000) circle (0.200000) node [label=center:{$\scriptstyle {3} $}]{};
\filldraw[fill=red, draw=black] (1.500000,2.250000) circle (0.200000) node [label=center:{$\scriptstyle {4} $}]{};
\filldraw[fill=white, draw=black] (0.000000,0.750000) circle (0.200000) node [label=center:{$\scriptstyle {5} $}]{};
\filldraw[fill=white, draw=black] (1.500000,0.750000) circle (0.200000) node [label=center:{$\scriptstyle {6} $}]{};
\filldraw[fill=white, draw=black] (0.750000,3.000000) circle (0.200000) node [label=center:{$\scriptstyle {7} $}]{};
\filldraw[fill=white, draw=black] (2.250000,3.000000) circle (0.200000) node [label=center:{$\scriptstyle {8} $}]{};
\filldraw[fill=white, draw=black] (0.750000,1.500000) circle (0.200000) node [label=center:{$\scriptstyle {9} $}]{};
\filldraw[fill=white, draw=black] (2.250000,1.500000) circle (0.200000) node [label=center:{$\scriptstyle {10} $}]{};
\filldraw[fill=white, draw=black] (0.750000,0.000000) circle (0.200000) node [label=center:{$\scriptstyle {11} $}]{};
\filldraw[fill=white, draw=black] (2.250000,0.000000) circle (0.200000) node [label=center:{$\scriptstyle {12} $}]{};
\filldraw[fill=white, draw=black] (0.550000,3.550000) rectangle (0.950000,3.950000) node [label=above right:{}]{};\filldraw[fill=white, draw=white] (0.750000,3.750000) circle (0.100000) node [label=center:{$\scriptstyle {} $}]{};
\filldraw[fill=white, draw=black] (2.050000,3.550000) rectangle (2.450000,3.950000) node [label=above right:{}]{};\filldraw[fill=white, draw=white] (2.250000,3.750000) circle (0.100000) node [label=center:{$\scriptstyle {} $}]{};
\filldraw[fill=white, draw=black] (0.550000,2.050000) rectangle (0.950000,2.450000) node [label=above right:{}]{};\filldraw[fill=white, draw=white] (0.750000,2.250000) circle (0.100000) node [label=center:{$\scriptstyle {} $}]{};
\filldraw[fill=white, draw=black] (2.050000,2.050000) rectangle (2.450000,2.450000) node [label=above right:{}]{};\filldraw[fill=white, draw=white] (2.250000,2.250000) circle (0.100000) node [label=center:{$\scriptstyle {} $}]{};
\filldraw[fill=white, draw=black] (0.550000,0.550000) rectangle (0.950000,0.950000) node [label=above right:{}]{};\filldraw[fill=white, draw=white] (0.750000,0.750000) circle (0.100000) node [label=center:{$\scriptstyle {} $}]{};
\filldraw[fill=white, draw=black] (2.050000,0.550000) rectangle (2.450000,0.950000) node [label=above right:{}]{};\filldraw[fill=white, draw=white] (2.250000,0.750000) circle (0.100000) node [label=center:{$\scriptstyle {} $}]{};
\filldraw[fill=white, draw=black] (-0.200000,2.800000) rectangle (0.200000,3.200000) node [label=above right:{}]{};\filldraw[fill=white, draw=white] (0.000000,3.000000) circle (0.100000) node [label=center:{$\scriptstyle {} $}]{};
\filldraw[fill=white, draw=black] (1.300000,2.800000) rectangle (1.700000,3.200000) node [label=above right:{}]{};\filldraw[fill=white, draw=white] (1.500000,3.000000) circle (0.100000) node [label=center:{$\scriptstyle {} $}]{};
\filldraw[fill=white, draw=black] (-0.200000,1.300000) rectangle (0.200000,1.700000) node [label=above right:{}]{};\filldraw[fill=white, draw=white] (0.000000,1.500000) circle (0.100000) node [label=center:{$\scriptstyle {} $}]{};
\filldraw[fill=white, draw=black] (1.300000,1.300000) rectangle (1.700000,1.700000) node [label=above right:{}]{};\filldraw[fill=white, draw=white] (1.500000,1.500000) circle (0.100000) node [label=center:{$\scriptstyle {} $}]{};
\filldraw[fill=white, draw=black] (-0.200000,-0.200000) rectangle (0.200000,0.200000) node [label=above right:{}]{};\filldraw[fill=white, draw=white] (0.000000,0.000000) circle (0.100000) node [label=center:{$\scriptstyle {} $}]{};
\filldraw[fill=white, draw=black] (1.300000,-0.200000) rectangle (1.700000,0.200000) node [label=above right:{}]{};\filldraw[fill=white, draw=white] (1.500000,0.000000) circle (0.100000) node [label=center:{$\scriptstyle {} $}]{};
\end{tikzpicture}
    \caption{The $H_Z$ sub-component of the $[[12,2,2]]$ XZZX toric code depicted in Fig~\ref{fig:xzzx_toric_hz_notwist}. The Hadamard rotation ensures that the $H_Z$ component decouples to a set of closed-loop repetition codes.}
    \label{fig:xzzx_toric_hz_notwist}
\end{figure}
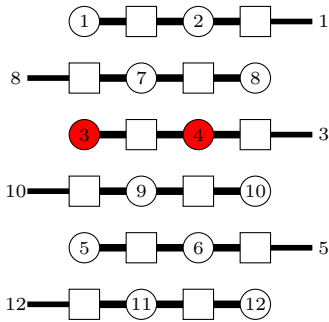

In addition to transforming the stabilisers, the Hadmard rotation also transforms the logical operators. For any CSS hypergraph product code, a basis of logical operators can be written in the form
\begin{equation}\label{eq:l_css}
	L_{\rm CSS}=\left[\begin{array}{@{} cc|cc @{}}
		0&0&L_{Z1}&L_{Z2}\\
		L_{X1}&L_{X2}&0&0
	\end{array}\right]\rm,
\end{equation} 
where we again separate the operators into the components that act on the sector one qubits and sector two qubits. In addition to transforming the code stabilisers, the Hadmard rotation also transforms the logical operators so that they read
\begin{equation}\label{eq:l_css}
	L_{\rm XZZX}=\left[\begin{array}{@{} cc|cc @{}}
		0&L_{Z2}&L_{Z1}&0\\
		L_{X1}&0&0&L_{X2}
	\end{array}\right]\rm.
\end{equation} 
From the above, it follows that the Hadamard rotation does not alter the weight of any logical operator and leaves the code parameters unchanged. Consequently, we expect the depolarising noise threshold of the XZZX code to be the same as that for the CSS toric code. However, under biased noise, the performance of the XZZX toric code immediately begins to outperform that of the CSS toric code. To understand this, we again consider the case of infinite $X$-bias, $\eta_X=\infty$, where the performance of the code is dependent only on the sub-component $H_Z$. This sub-component is depicted for the $[[12,2,2]]$ toric code in Fig.~\ref{fig:xzzx_toric_hz_notwist}, where we see that it now consists of a set of decoupled closed-loop repetition codes. The threshold of the repetition code is upper-bounded at $50\%$. As a result, the upper-bound on the threshold of the XZZX toric code grows from $18.9\%$ to $50\%$ with increasing noise bias. Herein lies the principal benefit of the XZZX construction: the Hadamard rotation modifies the stabilisers so that the $H_X$ and $H_Z$ components are built from copies of the seed codes (and their transposes). Therefore, provided the seeds codes are themselves \textit{good} classical codes, the resultant XZZX hypergraph product will also be a good code in the limit of infinite-bias. In the case of the XZZX code, the seeds are repetition codes and the resultant infinite-bias threshold saturates the Hashing bound.

\subsection{The XZZX toric code with twisted boundary conditions}

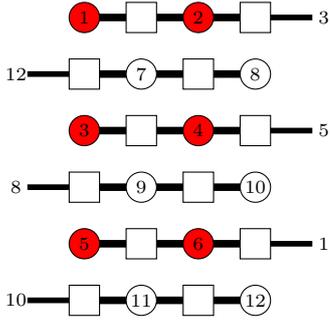
\begin{figure}
    \centering
    \begin{tikzpicture}
\draw[black,thick,line width=1.00mm] (0.0,3.75) -- (0.75,3.75);
\draw[black,thick,line width=1.00mm] (1.5,3.75) -- (0.75,3.75);
\draw[black,thick,line width=1.00mm] (1.5,3.75) -- (2.25,3.75);
\draw[black,thick,line width=0.75mm] (3.0,3.75) -- (2.25,3.75);
\node () at (3.150000,3.750000)  [] {\scriptsize 3};
\draw[black,thick,line width=1.00mm] (0.0,2.25) -- (0.75,2.25);
\draw[black,thick,line width=1.00mm] (1.5,2.25) -- (0.75,2.25);
\draw[black,thick,line width=1.00mm] (1.5,2.25) -- (2.25,2.25);
\draw[black,thick,line width=0.75mm] (3.0,2.25) -- (2.25,2.25);
\node () at (3.150000,2.250000)  [] {\scriptsize 5};
\draw[black,thick,line width=1.00mm] (0.0,0.75) -- (0.75,0.75);
\draw[black,thick,line width=1.00mm] (1.5,0.75) -- (0.75,0.75);
\draw[black,thick,line width=0.75mm] (3.0,0.75) -- (2.25,0.75);
\node () at (3.150000,0.750000)  [] {\scriptsize 1};
\draw[black,thick,line width=1.00mm] (1.5,0.75) -- (2.25,0.75);
\draw[black,thick,line width=1.00mm] (0.75,3.0) -- (0.0,3.0);
\draw[black,thick,line width=0.75mm] (-0.75,3.0) -- (0.0,3.0);
\node () at (-0.900000,3.000000)  [] {\scriptsize 12};
\draw[black,thick,line width=1.00mm] (0.75,3.0) -- (1.5,3.0);
\draw[black,thick,line width=1.00mm] (2.25,3.0) -- (1.5,3.0);
\draw[black,thick,line width=0.75mm] (-0.75,1.5) -- (0.0,1.5);
\node () at (-0.900000,1.500000)  [] {\scriptsize 8};
\draw[black,thick,line width=1.00mm] (0.75,1.5) -- (0.0,1.5);
\draw[black,thick,line width=1.00mm] (0.75,1.5) -- (1.5,1.5);
\draw[black,thick,line width=1.00mm] (2.25,1.5) -- (1.5,1.5);
\draw[black,thick,line width=0.75mm] (-0.75,0.0) -- (0.0,0.0);
\node () at (-0.900000,0.000000)  [] {\scriptsize 10};
\draw[black,thick,line width=1.00mm] (0.75,0.0) -- (0.0,0.0);
\draw[black,thick,line width=1.00mm] (0.75,0.0) -- (1.5,0.0);
\draw[black,thick,line width=1.00mm] (2.25,0.0) -- (1.5,0.0);
\filldraw[fill=red, draw=black] (0.000000,3.750000) circle (0.200000) node [label=center:{$\scriptstyle {1} $}]{};
\filldraw[fill=red, draw=black] (1.500000,3.750000) circle (0.200000) node [label=center:{$\scriptstyle {2} $}]{};
\filldraw[fill=red, draw=black] (0.000000,2.250000) circle (0.200000) node [label=center:{$\scriptstyle {3} $}]{};
\filldraw[fill=red, draw=black] (1.500000,2.250000) circle (0.200000) node [label=center:{$\scriptstyle {4} $}]{};
\filldraw[fill=red, draw=black] (0.000000,0.750000) circle (0.200000) node [label=center:{$\scriptstyle {5} $}]{};
\filldraw[fill=red, draw=black] (1.500000,0.750000) circle (0.200000) node [label=center:{$\scriptstyle {6} $}]{};
\filldraw[fill=white, draw=black] (0.750000,3.000000) circle (0.200000) node [label=center:{$\scriptstyle {7} $}]{};
\filldraw[fill=white, draw=black] (2.250000,3.000000) circle (0.200000) node [label=center:{$\scriptstyle {8} $}]{};
\filldraw[fill=white, draw=black] (0.750000,1.500000) circle (0.200000) node [label=center:{$\scriptstyle {9} $}]{};
\filldraw[fill=white, draw=black] (2.250000,1.500000) circle (0.200000) node [label=center:{$\scriptstyle {10} $}]{};
\filldraw[fill=white, draw=black] (0.750000,0.000000) circle (0.200000) node [label=center:{$\scriptstyle {11} $}]{};
\filldraw[fill=white, draw=black] (2.250000,0.000000) circle (0.200000) node [label=center:{$\scriptstyle {12} $}]{};
\filldraw[fill=white, draw=black] (0.550000,3.550000) rectangle (0.950000,3.950000) node [label=above right:{}]{};\filldraw[fill=white, draw=white] (0.750000,3.750000) circle (0.100000) node [label=center:{$\scriptstyle {} $}]{};
\filldraw[fill=white, draw=black] (2.050000,3.550000) rectangle (2.450000,3.950000) node [label=above right:{}]{};\filldraw[fill=white, draw=white] (2.250000,3.750000) circle (0.100000) node [label=center:{$\scriptstyle {} $}]{};
\filldraw[fill=white, draw=black] (0.550000,2.050000) rectangle (0.950000,2.450000) node [label=above right:{}]{};\filldraw[fill=white, draw=white] (0.750000,2.250000) circle (0.100000) node [label=center:{$\scriptstyle {} $}]{};
\filldraw[fill=white, draw=black] (2.050000,2.050000) rectangle (2.450000,2.450000) node [label=above right:{}]{};\filldraw[fill=white, draw=white] (2.250000,2.250000) circle (0.100000) node [label=center:{$\scriptstyle {} $}]{};
\filldraw[fill=white, draw=black] (0.550000,0.550000) rectangle (0.950000,0.950000) node [label=above right:{}]{};\filldraw[fill=white, draw=white] (0.750000,0.750000) circle (0.100000) node [label=center:{$\scriptstyle {} $}]{};
\filldraw[fill=white, draw=black] (2.050000,0.550000) rectangle (2.450000,0.950000) node [label=above right:{}]{};\filldraw[fill=white, draw=white] (2.250000,0.750000) circle (0.100000) node [label=center:{$\scriptstyle {} $}]{};
\filldraw[fill=white, draw=black] (-0.200000,2.800000) rectangle (0.200000,3.200000) node [label=above right:{}]{};\filldraw[fill=white, draw=white] (0.000000,3.000000) circle (0.100000) node [label=center:{$\scriptstyle {} $}]{};
\filldraw[fill=white, draw=black] (1.300000,2.800000) rectangle (1.700000,3.200000) node [label=above right:{}]{};\filldraw[fill=white, draw=white] (1.500000,3.000000) circle (0.100000) node [label=center:{$\scriptstyle {} $}]{};
\filldraw[fill=white, draw=black] (-0.200000,1.300000) rectangle (0.200000,1.700000) node [label=above right:{}]{};\filldraw[fill=white, draw=white] (0.000000,1.500000) circle (0.100000) node [label=center:{$\scriptstyle {} $}]{};
\filldraw[fill=white, draw=black] (1.300000,1.300000) rectangle (1.700000,1.700000) node [label=above right:{}]{};\filldraw[fill=white, draw=white] (1.500000,1.500000) circle (0.100000) node [label=center:{$\scriptstyle {} $}]{};
\filldraw[fill=white, draw=black] (-0.200000,-0.200000) rectangle (0.200000,0.200000) node [label=above right:{}]{};\filldraw[fill=white, draw=white] (0.000000,0.000000) circle (0.100000) node [label=center:{$\scriptstyle {} $}]{};
\filldraw[fill=white, draw=black] (1.300000,-0.200000) rectangle (1.700000,0.200000) node [label=above right:{}]{};\filldraw[fill=white, draw=white] (1.500000,0.000000) circle (0.100000) node [label=center:{$\scriptstyle {} $}]{};
\end{tikzpicture}
    \caption{The $H_X$ sub-component of the twisted $[[12,2,3]]$ XZZX toric code. Following the boundary twist, the minimum-weight of a non-trivial logical operator in $H_Z$ has distance six. The qubits filled in red show an example of such a logical operator that spans across the sector one qubits.}
    \label{fig:xzzx_toric_hz}
\end{figure}

In the infinite $X$-bias regime, the $[[12,2,2]]$ XZZX toric code has logical-$X$ distance $d_X=2$. The nodes filled in red Fig.~\ref{fig:xzzx_toric_hz_notwist} show an example of a minimum-weight logical operator that spans the lattice. One way of improving the infinite-bias distance $d_X$ is to increase the width of the lattice: this amounts to using a longer repetition code as the seed $H_2$. The disadvantage of this method is that it results in more qubits and a reduction in the code rate. An alternative, more qubit-efficient technique for improving $d_X$, is to modify the code via a \textit{boundary twist}. In the CSS toric code derived from the hypergraph product, boundary checks connect to qubits on the opposite side of the lattice. Fig.~\ref{fig:xzzx_toric} shows an XZZX toric code where the boundary conditions for the $H_Z$ component have been \textit{twisted} so that each boundary check connects instead to the qubit shifted one row below on the opposite side of the lattice. The resultant code has an improved distance of $d=3$ giving it parameters $[[12,2,3]]$. In the infinite-bias regime, the distance is increased even further. Fig.~\ref{fig:xzzx_toric_hz} shows a logical operator (red nodes) on the $H_Z$ sub-component of the twisted code. We see that this logical operator has weight six and covers all the qubits in sector one. Similarly, a second independent logical operator can be defined that covers all sector two qubits. In the limit of infinite $X$-bias, the $X$-distance is therefore $d_X=6$. 

\jr{In general, we define an XZZX twisted toric code on an $n_1\times n_2$ lattice where $n_1$ corresponds to the number of data qubits spanning the lattice vertically and $n_2$ the number of data qubits spanning the lattice horizontally. Eg. the lattice depicted in Fig.~\ref{fig:xzzx_toric} has lattice parameters $n_1=3$ and $n_2=2$. The parity check matrix of the XZZX twisted toric code is written
\begin{equation}\label{eq:hgp_xzzx_twist}
\arraycolsep=4pt\def\arraystretch{1}
\thickmuskip=0.5\thickmuskip
\begin{split}
    &H=\left[H_X|H_Z\right]\\
    &=\left[\begin{array}{@{} cc|cc @{}}
    0 & H_1^T{\otimes} \openone_{n_2} & H^{\rm rep}_{N/2} & 0 \\
    H_1{\otimes}\openone_{n_2} & 0 & 0 & (H^{\rm rep}_{N/2})^T
    \end{array}
    \right]\rm,
    \end{split}
\end{equation}
where $N$ is the qubit block length $N=2n_1n_2$. We see that the above differs from Eq.~(\ref{eq:hgp_xzzx}) in that the $H_Z$ component has been rewritten in terms of two repetition codes of length $N/2$.} The first repetition codes wraps around the sector one qubits and the second around the sector two qubits. Note that for the example in this section, we have twisted only the $H_Z$ component. An equivalent XZZX code could be constructed by instead twisting the $H_X$ component.

In the CSS setting, boundary twists have previously been used to improve code performance by Kovalev and Pryadko \cite{kovalev12}. Boundary twists have also served a role in the recent discoveries of high-performance quantum LDPC codes \cite{hastings2021fiber,breuckmann21}.

\subsection{Lifted product representation of the XZZX toric code}

Following the boundary twist, the parity check matrix of the XZZX twisted toric code given by Eq.~(\ref{eq:hgp_xzzx_twist}) can no longer be derived from the hypergraph product. In this section, we show that the XZZX twisted toric code is in fact an instance of a lifted product code.

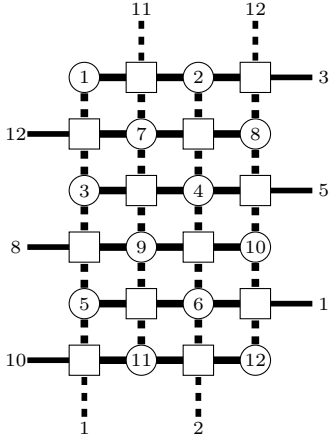
\begin{figure}
    \centering
    \begin{tikzpicture}
\draw[black,thick,dashed,line width=1.00mm] (0.75,3.0) -- (0.75,3.75);
\draw[black,thick,dashed,line width=0.75mm] (0.75,4.5) -- (0.75,3.75);
\node () at (0.750000,4.650000)  [] {\scriptsize 11};
\draw[black,thick,dashed,line width=1.00mm] (2.25,3.0) -- (2.25,3.75);
\draw[black,thick,dashed,line width=0.75mm] (2.25,4.5) -- (2.25,3.75);
\node () at (2.250000,4.650000)  [] {\scriptsize 12};
\draw[black,thick,dashed,line width=1.00mm] (0.75,3.0) -- (0.75,2.25);
\draw[black,thick,dashed,line width=1.00mm] (0.75,1.5) -- (0.75,2.25);
\draw[black,thick,dashed,line width=1.00mm] (2.25,3.0) -- (2.25,2.25);
\draw[black,thick,dashed,line width=1.00mm] (2.25,1.5) -- (2.25,2.25);
\draw[black,thick,dashed,line width=1.00mm] (0.75,1.5) -- (0.75,0.75);
\draw[black,thick,dashed,line width=1.00mm] (0.75,0.0) -- (0.75,0.75);
\draw[black,thick,dashed,line width=1.00mm] (2.25,1.5) -- (2.25,0.75);
\draw[black,thick,dashed,line width=1.00mm] (2.25,0.0) -- (2.25,0.75);
\draw[black,thick,dashed,line width=1.00mm] (0.0,3.75) -- (0.0,3.0);
\draw[black,thick,dashed,line width=1.00mm] (0.0,2.25) -- (0.0,3.0);
\draw[black,thick,dashed,line width=1.00mm] (1.5,3.75) -- (1.5,3.0);
\draw[black,thick,dashed,line width=1.00mm] (1.5,2.25) -- (1.5,3.0);
\draw[black,thick,dashed,line width=1.00mm] (0.0,2.25) -- (0.0,1.5);
\draw[black,thick,dashed,line width=1.00mm] (0.0,0.75) -- (0.0,1.5);
\draw[black,thick,dashed,line width=1.00mm] (1.5,2.25) -- (1.5,1.5);
\draw[black,thick,dashed,line width=1.00mm] (1.5,0.75) -- (1.5,1.5);
\draw[black,thick,dashed,line width=0.75mm] (0.0,-0.75) -- (0.0,0.0);
\node () at (0.000000,-0.900000)  [] {\scriptsize 1};
\draw[black,thick,dashed,line width=1.00mm] (0.0,0.75) -- (0.0,0.0);
\draw[black,thick,dashed,line width=0.75mm] (1.5,-0.75) -- (1.5,0.0);
\node () at (1.500000,-0.900000)  [] {\scriptsize 2};
\draw[black,thick,dashed,line width=1.00mm] (1.5,0.75) -- (1.5,0.0);
\draw[black,thick,line width=1.00mm] (0.0,3.75) -- (0.75,3.75);
\draw[black,thick,line width=1.00mm] (1.5,3.75) -- (0.75,3.75);
\draw[black,thick,line width=1.00mm] (1.5,3.75) -- (2.25,3.75);
\draw[black,thick,line width=0.75mm] (3.0,3.75) -- (2.25,3.75);
\node () at (3.150000,3.750000)  [] {\scriptsize 3};
\draw[black,thick,line width=1.00mm] (0.0,2.25) -- (0.75,2.25);
\draw[black,thick,line width=1.00mm] (1.5,2.25) -- (0.75,2.25);
\draw[black,thick,line width=1.00mm] (1.5,2.25) -- (2.25,2.25);
\draw[black,thick,line width=0.75mm] (3.0,2.25) -- (2.25,2.25);
\node () at (3.150000,2.250000)  [] {\scriptsize 5};
\draw[black,thick,line width=1.00mm] (0.0,0.75) -- (0.75,0.75);
\draw[black,thick,line width=1.00mm] (1.5,0.75) -- (0.75,0.75);
\draw[black,thick,line width=0.75mm] (3.0,0.75) -- (2.25,0.75);
\node () at (3.150000,0.750000)  [] {\scriptsize 1};
\draw[black,thick,line width=1.00mm] (1.5,0.75) -- (2.25,0.75);
\draw[black,thick,line width=1.00mm] (0.75,3.0) -- (0.0,3.0);
\draw[black,thick,line width=0.75mm] (-0.75,3.0) -- (0.0,3.0);
\node () at (-0.900000,3.000000)  [] {\scriptsize 12};
\draw[black,thick,line width=1.00mm] (0.75,3.0) -- (1.5,3.0);
\draw[black,thick,line width=1.00mm] (2.25,3.0) -- (1.5,3.0);
\draw[black,thick,line width=0.75mm] (-0.75,1.5) -- (0.0,1.5);
\node () at (-0.900000,1.500000)  [] {\scriptsize 8};
\draw[black,thick,line width=1.00mm] (0.75,1.5) -- (0.0,1.5);
\draw[black,thick,line width=1.00mm] (0.75,1.5) -- (1.5,1.5);
\draw[black,thick,line width=1.00mm] (2.25,1.5) -- (1.5,1.5);
\draw[black,thick,line width=0.75mm] (-0.75,0.0) -- (0.0,0.0);
\node () at (-0.900000,0.000000)  [] {\scriptsize 10};
\draw[black,thick,line width=1.00mm] (0.75,0.0) -- (0.0,0.0);
\draw[black,thick,line width=1.00mm] (0.75,0.0) -- (1.5,0.0);
\draw[black,thick,line width=1.00mm] (2.25,0.0) -- (1.5,0.0);
\filldraw[fill=white, draw=black] (0.000000,3.750000) circle (0.200000) node [label=center:{$\scriptstyle {1} $}]{};
\filldraw[fill=white, draw=black] (1.500000,3.750000) circle (0.200000) node [label=center:{$\scriptstyle {2} $}]{};
\filldraw[fill=white, draw=black] (0.000000,2.250000) circle (0.200000) node [label=center:{$\scriptstyle {3} $}]{};
\filldraw[fill=white, draw=black] (1.500000,2.250000) circle (0.200000) node [label=center:{$\scriptstyle {4} $}]{};
\filldraw[fill=white, draw=black] (0.000000,0.750000) circle (0.200000) node [label=center:{$\scriptstyle {5} $}]{};
\filldraw[fill=white, draw=black] (1.500000,0.750000) circle (0.200000) node [label=center:{$\scriptstyle {6} $}]{};
\filldraw[fill=white, draw=black] (0.750000,3.000000) circle (0.200000) node [label=center:{$\scriptstyle {7} $}]{};
\filldraw[fill=white, draw=black] (2.250000,3.000000) circle (0.200000) node [label=center:{$\scriptstyle {8} $}]{};
\filldraw[fill=white, draw=black] (0.750000,1.500000) circle (0.200000) node [label=center:{$\scriptstyle {9} $}]{};
\filldraw[fill=white, draw=black] (2.250000,1.500000) circle (0.200000) node [label=center:{$\scriptstyle {10} $}]{};
\filldraw[fill=white, draw=black] (0.750000,0.000000) circle (0.200000) node [label=center:{$\scriptstyle {11} $}]{};
\filldraw[fill=white, draw=black] (2.250000,0.000000) circle (0.200000) node [label=center:{$\scriptstyle {12} $}]{};
\filldraw[fill=white, draw=black] (0.550000,3.550000) rectangle (0.950000,3.950000) node [label=above right:{}]{};\filldraw[fill=white, draw=white] (0.750000,3.750000) circle (0.100000) node [label=center:{$\scriptstyle {} $}]{};
\filldraw[fill=white, draw=black] (2.050000,3.550000) rectangle (2.450000,3.950000) node [label=above right:{}]{};\filldraw[fill=white, draw=white] (2.250000,3.750000) circle (0.100000) node [label=center:{$\scriptstyle {} $}]{};
\filldraw[fill=white, draw=black] (0.550000,2.050000) rectangle (0.950000,2.450000) node [label=above right:{}]{};\filldraw[fill=white, draw=white] (0.750000,2.250000) circle (0.100000) node [label=center:{$\scriptstyle {} $}]{};
\filldraw[fill=white, draw=black] (2.050000,2.050000) rectangle (2.450000,2.450000) node [label=above right:{}]{};\filldraw[fill=white, draw=white] (2.250000,2.250000) circle (0.100000) node [label=center:{$\scriptstyle {} $}]{};
\filldraw[fill=white, draw=black] (0.550000,0.550000) rectangle (0.950000,0.950000) node [label=above right:{}]{};\filldraw[fill=white, draw=white] (0.750000,0.750000) circle (0.100000) node [label=center:{$\scriptstyle {} $}]{};
\filldraw[fill=white, draw=black] (2.050000,0.550000) rectangle (2.450000,0.950000) node [label=above right:{}]{};\filldraw[fill=white, draw=white] (2.250000,0.750000) circle (0.100000) node [label=center:{$\scriptstyle {} $}]{};
\filldraw[fill=white, draw=black] (-0.200000,2.800000) rectangle (0.200000,3.200000) node [label=above right:{}]{};\filldraw[fill=white, draw=white] (0.000000,3.000000) circle (0.100000) node [label=center:{$\scriptstyle {} $}]{};
\filldraw[fill=white, draw=black] (1.300000,2.800000) rectangle (1.700000,3.200000) node [label=above right:{}]{};\filldraw[fill=white, draw=white] (1.500000,3.000000) circle (0.100000) node [label=center:{$\scriptstyle {} $}]{};
\filldraw[fill=white, draw=black] (-0.200000,1.300000) rectangle (0.200000,1.700000) node [label=above right:{}]{};\filldraw[fill=white, draw=white] (0.000000,1.500000) circle (0.100000) node [label=center:{$\scriptstyle {} $}]{};
\filldraw[fill=white, draw=black] (1.300000,1.300000) rectangle (1.700000,1.700000) node [label=above right:{}]{};\filldraw[fill=white, draw=white] (1.500000,1.500000) circle (0.100000) node [label=center:{$\scriptstyle {} $}]{};
\filldraw[fill=white, draw=black] (-0.200000,-0.200000) rectangle (0.200000,0.200000) node [label=above right:{}]{};\filldraw[fill=white, draw=white] (0.000000,0.000000) circle (0.100000) node [label=center:{$\scriptstyle {} $}]{};
\filldraw[fill=white, draw=black] (1.300000,-0.200000) rectangle (1.700000,0.200000) node [label=above right:{}]{};\filldraw[fill=white, draw=white] (1.500000,0.000000) circle (0.100000) node [label=center:{$\scriptstyle {} $}]{};
\end{tikzpicture}
    \caption{The $[[12,2,3]]$ XZZX twisted toric code. The boundary edges have been twisted so that each boundary check node now connects to the qubit shifted one position below on the opposite side of the lattice. This increases the length of the non-trivial logical operators, and results in an improvement in the minimum distance of the code from $d=2$ to $d=3$.}
    \label{fig:xzzx_toric}
\end{figure}

In order to recast the XZZX twisted code as a lifted product we need to find a protograph representation of the parity check matrix in Eq.~ (\ref{eq:hgp_xzzx_twist}). From Example~\ref{ex:proto_rep}, we note that the parity check matrix of the repetition code $H^{\rm rep}_{N/2}$ can be expressed in terms of the following protograph
\begin{equation}
    H^{\rm rep}_{N/2}=\mathfrak{B}([\lambda^0_{N/2}+\lambda^1_{N/2}])\rm.
\end{equation}
\jr{Here $N=2n_1n_2$ where $n_1$ and $n_2$ are the vertical and horizontal widths of the lattice respectively. By inspection of the factor graph in Fig.~\ref{fig:xzzx_toric}, we see that the $X$-parity checks connect qubit $i$ to qubit $(i+n_2) \ {\rm mod} \ N/2$.} Using this periodicity, we find that
\begin{equation}
H_1\otimes \openone_{n_2}=\mathfrak{B}([
\lambda_{N/2}^0+\lambda_{N/2}^{n_2}])\rm.    
\end{equation}
The protograph representation of Eq.~(\ref{eq:hgp_xzzx_twist}) is then given by
\begin{equation}\label{eq:toric_proto}
\arraycolsep=4pt\def\arraystretch{1}
\thickmuskip=0.5\thickmuskip
\begin{split}
    A=\left[A_X|A_Z\right]  =\left[\begin{array}{@{} cc|cc @{}}
    0 & A_1^T & A_2 & 0 \\
    A_1 & 0 & 0 & A_2^T
    \end{array}
    \right]\rm,
    \end{split}
\end{equation}
where $A_1=[\lambda_{N/2}^0+\lambda_{N/2}^{n_2}]$ and $A_2=[
\lambda_{N/2}^0+\lambda_{N/2}^{1}]$. \jr{We now note that  $A_k=A_k\otimes E_1=E_1 \otimes A_k$, where where $E_1$ is a $1\times1$ instance of the identity protograph defined in Eq.~(\ref{eq:protograph_identity}). Using this relation, we can rewrite Eq.~(\ref{eq:toric_proto}) as follows
\begin{equation}\label{eq:toric_lifted}
\arraycolsep=3pt\def\arraystretch{1}
\thickmuskip=0.5\thickmuskip
\begin{split}
    A=\left[\begin{array}{@{} cc|cc @{}}
    0 & A_1^T\otimes E_1 & E_1\otimes A_2 & 0 \\
    A_1\otimes E_1 & 0 & 0 & E_1\otimes A_2^T
    \end{array}
    \right]\rm.
    \end{split}
\end{equation}
Now, note the similarity of the above to the lifted product defined in Eq.~(\ref{eq:lifted_hgp}). The protograph for the XZZX toric code, Eq.~(\ref{eq:toric_lifted}), is simply the lifted product of the $1{\times}1$ seed protographs, $A_1$ and $A_2$, followed by Hadamard rotation on the second block of $N/2$ qubits. It is clear that this form of the lifted product encapsulates both the stabiliser redefinition and boundary twists that give the XZZX twisted toric code such high performance under biased noise. 
}

\subsection{The bias-tailored lifted product}

In the previous section we saw that the XZZX twisted toric code can be described as a special case of the lifted product applied to two $1{\times}1$ protographs. However, we need not be restricted to this special case: any combination of seed protographs $A_1$ and $A_2$ can be used to construct quantum LDPC codes. Given two seed protographs,
$A_1\in\mathbb{A}_L^{m'_1{\times}n'_1}$ with dimensions $m'_1{\times}n'_1$ and $A_2\in\mathbb{A}_L^{m'_2{\times}n'_2}$ with dimensions $m'_2{\times}n'_2$, we can define a bias-tailored lifted product code with the following protograph
\begin{equation}\label{eq:proto_ldpc_bias_tailored}
\arraycolsep=4pt\def\arraystretch{1}
\thickmuskip=0.5\thickmuskip
\begin{split}
    &A_{Q}=\left[A_X|A_Z\right]\\
    &=\left[\begin{array}{@{} cc|cc @{}}
    0 & (A_1)^T{\otimes} E_{m'_2} & E_{n'_1}{\otimes} A_2 & 0 \\
    A_1{\otimes}E_{n'_2} & 0 & 0 & E_{m'_1}{\otimes} (A_2)^T
    \end{array}
    \right]\rm.
    \end{split}
\end{equation}
\jr{where $E_m$ is a $m\times m$ instance of the identity protograph defined in Eq.~(\ref{eq:protograph_identity}).} The above defines a non-CSS code with a parity check matrix given by $\mathfrak{B}(A_Q)$. 

\begin{example}\label{example:lpc1}
We now consider a bias-tailored lifted product where both $A_1$ and $A_2$ are the $4{\times}4$ protograph $A$ given by Eq.~(\ref{eq:proto1}). Only a Hadamard rotation differentiates the bias-tailored lifted product (see Eq.~(\ref{eq:proto_ldpc_bias_tailored})) from the standard lifted product (see Eq.(\ref{eq:lifted_hgp})). As such, the biased-tailored code with $L=13$ has the same parameters $[[416,18,d\leq20]]$ as the lifted product code from Example \ref{example:lpc}. However, under infinite $X$-bias, only the $A_Z$ component of the protograph is relevant. This is written explicitly for the $[[416,18,d\leq20]]$ code below
\begin{equation}\label{eq:lpc1_decoupled}
\arraycolsep=5pt\def\arraystretch{1}
\thickmuskip=0.5\thickmuskip
A_Z=\left[\begin{array}{@{} cccccccc @{}}
    A   &   0   &   0  &   0   &   0   &   0   &   0   &   0   \\
    0   &   A   &   0   &   0  &   0   &   0   &   0   &   0   \\
    0   &   0   &   A   &   0   &   0  & 0   &   0   &   0   \\
    0   &   0   &   0   &   A   &   0   &   0   &   0   &   0   \\
    0   &   0   &   0   &   0   &   A^T   &   0   &   0   &   0   \\
    0   &   0   &   0   &   0   &   0   &  A^T  &   0   &   0   \\
    0   &   0   &   0   &   0   &   0   &   0   &   A^T   &   0   \\
    0   &   0   &   0   &   0   &   0   &   0   &   0   & A^T
    \end{array}\right]\rm.
\end{equation}
Here we see that the $A_Z$ protograph consists of a set of decoupled copies of the the seed protograph $A$ and its transpose. Both the code $\mathfrak{B}(A)$ and its transpose $\mathfrak{B}(A^T)$ have parameters $[52,3,26]$. Consequently the infinate bias $X$-distance is $d_X=26$. This compares favourably to the distance of $d\leq20$ for depolarising noise.

\end{example}

\section{Numerical results}\label{sec:numerics}

To assess the performance of bias-tailored quantum LDPC codes we numerically estimate the logical error rates under various bias regimes. To this end, we run Monte Carlo simulations to randomly generate errors weighted according to an instance of the biased noise channel defined in Eq.~(\ref{eq:error_channel}). These errors are then corrected using the recovery operation output by a quantum decoder. In the following, we assume we are decoding an $N$-qubit quantum code, with parity check matrix $H\in\mathbb{F}_2^{M{\times}2N}$. The key steps in a single run of our simulation are summarised below
\begin{enumerate}
    \item Sample a Pauli error $\textbf{e}\in\mathbb{F}_2^{2N}$ from a distribution weighted by the chosen error channel's $p_X$, $p_Y$ and $p_Z$ probabilities.
    \item Calculate the error syndrome $\textbf{s}\in\mathbb{F}_2^{M}$ using Eq.~(\ref{eq:qsynd}).
    \item Decode the syndrome to obtain a recovery vector $\textbf{r}\in\mathbb{F}_2^{2N}$.
    \item Calculate the residual error $\textbf{r}'=\textbf{e}+\textbf{r}$.
    \item Check if $\textbf{r}'$ is a stabiliser. This amounts to verifying that $\textbf{r}'\in {\textsc{rowspace}}(H)$. If $\textbf{r}'$ is a stabiliser then the decoding is successful. If not, then $\textbf{r}'$ is a logical operator meaning the decoding has failed.
\end{enumerate}
The fraction of failed runs gives the simulated \textit{block error rate} $P_L$ of the quantum code. The block error rate tells us the total probability of an error occurring on any of the logical qubits. However, it does not take into account the number of logical qubits encoded by the code. A more useful metric for quantifying code performance is the \textit{word error rate} $P_W$  defined as
\begin{equation}
    P_W=1-(1-P_L)^{1/K}\rm,
\end{equation}
where $K$ is the number of encoded logical qubits. The word error rate can be thought of as the \textit{logical-error-rate-per-logical-qubit} and is particularly useful for comparing quantum LDPC codes of different sizes. For small logical error rates, the word error rate approximates to $P_W\approx P_L/K$.

\jr{When decoding CSS codes, a common strategy is to treat the $X$- and $Z$-stabilisers as two separate codes. The quantum code can then be decoded using two factor graphs of size $N$ instead of decoding the full quantum parity check matrix of size $2N$. Depending upon the complexity scaling of the decoding algorithm, this can result in considerable reductions in runtime. Bias-tailored lifted product codes are not CSS. However, their stabilisers are locally equivalent to a CSS code, differing only by the Hadamard rotation. Using this equivalence, we have found that the decoding graph can still be decoupled into two separate components of length $N$. A disadvantage of this simplification is that the resultant decoder ignores correlations between the two factor graphs that arise when a $Y$-error occurs. To counter this, we have developed a routine that implements a Bayesian update on the error channel of the second round of decoding conditional on the output of the first round of decoding. More details about our decoding procedures can be found in Appendix~\ref{app:simulation}.}

\subsection{Decoding methods} \label{sec:decoders}

 In our simulations we use one of two decoding methods: 1) minimum-weight perfect matching (MWPM) or 2) belief propagation plus ordered statistics decoding (BP+OSD). We now briefly describe each method.

\textbf{Minimum-weight perfect matching} (MWPM) is a decoding method that can be applied to any code where error chains produce pair-like syndromes \cite{Edmonds1965,Kolmogorov2009}. More concisely, MWPM requires any chain-like error $\textbf{e}$ to trigger a syndrome $\textbf{s}=H\cdot \textbf{e}$ of weight two such that ${\rm wt}(\textbf{s})=2$. On a factor graph, the triggered parity checks exist at the ends of the chain of errors. Given an ensemble of syndromes, the task of the MWPM matching decoder is to find the set of error chains (matchings) with the lowest total weight.  Examples of codes with parity check matrices that are amenable to MWPM decoding include: repetition codes, surface codes, toric codes, XZZX surface codes and XZZX toric codes. Note that at the boundary surface codes produce syndromes with weight one. However, this problem can be resolved by adding \textit{boundary nodes} to the factor graph and assigning their error probability to zero. In this work we use the \texttt{PyMatching} implementation of MWPM \cite{higgott2021pymatching} to decode instances of the CSS and XZZX toric code.

\textbf{Belief propagation plus ordered statistics decoding} (BP+OSD) is a versatile decoding method for quantum LDPC codes \cite{Panteleev_2019}. In contrast to MWPM, BP+OSD can decode codes which produce syndromes of weight greater than two. This makes it suitable for random codes constructed from the lifted product. The first step in BP+OSD is to apply the belief propagation (BP) algorithm, a standard technique for decoding classical LDPC codes \cite{kschischang2001factor,mackay1997near}. Belief propagation exploits the structure of a code's factor graph to efficiently compute the probability of each bit being errored given a measured syndrome. In general, BP can decode codes in time linear to the block length provided the factor graph is sufficiently sparse and has a low-density of short-length loops. For quantum codes, the latter requirement is problematic as quantum degeneracy necessarily leads to loop-like structures in the quantum factor graph. Consequently, the BP algorithm fails when degeneracy means there are multiple low-weight solutions for a certain syndrome. To counter this, the second stage of BP+OSD invokes a post-processing routine known as ordered statistics decoding (OSD) \cite{fossorier2001iterative}. The OSD step involves inverting the parity check matrix to yield a unique solution to the decoding problem thus removing any ambiguity due to quantum degeneracy. To ensure a low-weight solution is obtained, OSD uses the output of BP to choose a high-probability subset of qubits over which the inversion is performed. Numerical simulations have shown BP+OSD to be a highly-performant decoder for hypergraph product codes \cite{roffe2020decoding,Panteleev_2019}, lifted product \cite{Panteleev_2019} codes, 2D topological codes \cite{roffe2020decoding} and 3D topological codes \cite{quintavalle2021}. For this work, we make use of the BP+OSD implementation included in the \texttt{LDPC} package \cite{roffe_ldpc}.

\subsection{Toric codes with twisted boundaries}

\begin{figure}
    \centering
    \input{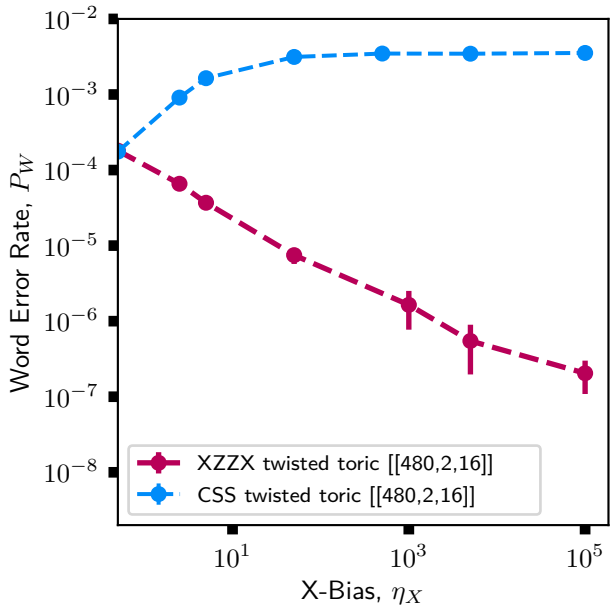}
    \caption{Decoding of the XZZX twisted toric code (red) vs. the CSS twisted toric code (blue) under increasing values of the $X$-bias. Both simulations were run using the MWPM decoder at a fixed physical qubit error rate of $p=0.06$. The $x$-axis intersects the $y$-axis at $\eta_X=0.5$ which corresponds to depolarising noise. The $X$-bias is increased by fixing $p_Y=p_Z$ and growing the value of $p_X$. Error bars show three times the standard deviation. }
    \label{fig:xzzx_css_comparison}
\end{figure}

We first present the finding of our numerical investigations of toric codes with twisted boundaries. As described in Section~\ref{sec:bias_tailored_qec}, such codes can be derived from the lifted product. We consider toric codes defined on a rectangular lattice with dimensions $n\times(n-1)$. These lattice parameters are chosen because they lead to the best code distances relative to the block length. For each instance of the $n\times (n-1)$ lattice, we construct two codes: 1) the twisted CSS toric code and 2) the XZZX twisted toric code. The twisted CSS toric code has parity check matrices derived from the lifted product defined in Eq.(\ref{eq:lifted_hgp}) with seed protographs $A_1=[\lambda_{N/2}^0+\lambda_{N/2}^{(n-1)}]$ and $A_2=[
\lambda_{N/2}^0+\lambda_{N/2}^{1}]$. Here, the block length is $N=2n(n-1)$. The XZZX twisted toric code is defined from the same protographs $A_1$ and $A_2$ using the bias-tailored lifted product defined in Eq.(\ref{eq:proto_ldpc_bias_tailored}). Fig.~\ref{fig:xzzx_toric} depicts an example of an XZZX twisted toric on a $3\times2$ qubit lattice.

Fig.~\ref{fig:xzzx_css_comparison} shows the MWPM decoding curves for a CSS and XZZX twisted toric code under different bias regimes for a fixed physical error rate of $p=0.06$. Each code is defined on a lattice with dimensions $16\times15$ giving code parameters $[[480,2,16]]$. The $X$-bias $\eta_X$ is plotted on the x-axis and the word error rate on the y-axis. For depolarising noise, the word error rate of the two codes is equal. With increasing bias, however, the decoding performance of the two codes immediately begins to differ. For the CSS code, the word error rate begins to rise, eventually converging to a value of approximately ten times the depolarising error rate. In contrast, increasing the bias for the XZZX version of the code leads to an exponential reduction in the word error rate. This effect can be attributed to the fact that XZZX codes decouple to independent copies of their seed codes in the high-bias regime. The full quantum code distance for depolarising noise is $d=16$, whereas the infinite bias distance for $X$-errors is $d_X=240$; an improvement in code performance is therefore expected as the bias is increased. For the standard CSS version of the code, no decoupling occurs, and the distance remains $d=16$ in all bias regimes.

\begin{figure}
    \centering
    \input{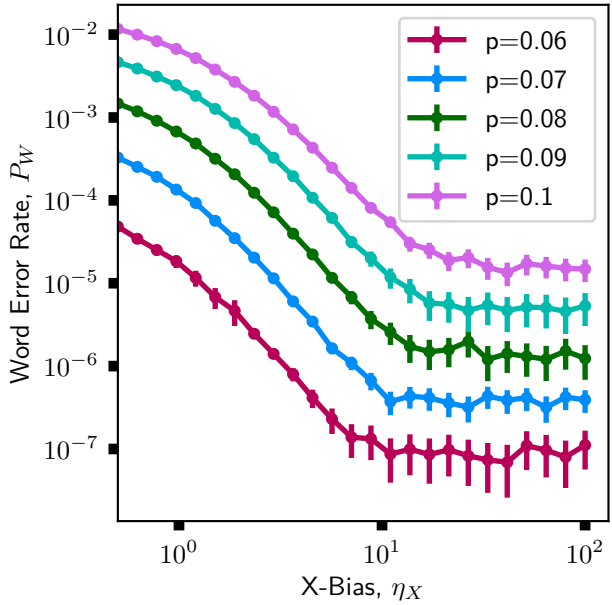}
    \caption{BP+OSD decoding of the $[[416,18,d\leq20]]$ bias-tailored lifted product code for physical error rates in the range $0.06\leq p \leq 0.10$. The error bars show three times the standard deviation.}
    \label{fig:lpc1}
\end{figure}

\subsection{Quantum LDPC codes}

We numerically investigate two quantum LDPC codes constructed from the bias-tailored lifted product. The first is the $[[416,18,d\leq 20]]$ code whose construction is outlined in Example~\ref{example:lpc1}. The second is a biased-tailored lifted product code  with parameters $[[882,24,d\leq24]]$. This code was first presented in CSS form in \cite{Panteleev_2019}, and the corresponding protographs can be found in Appendix~\ref{app:protographs}. For both codes, the performance is accessed by running simulations using the BP+OSD decoder. The lower bounds on the distances are estimated from the weights of the smallest observed logical operators during these simulations.

Fig.~\ref{fig:lpc1} shows the decoding plot of word error rate against $X$-bias for the $[[416,18,d\leq 20]]$ lifted product code for physical error rates in the range $0.06\leq p \leq 0.10$. As the $X$-bias $\eta_X$ is increased we initially see an exponential decrease in the error rate. However, the decoding curve plateaus for $\eta_X>10$. To understand why this levelling-off occurs we refer to Eq.~(\ref{eq:lpc1_decoupled}) which gives the quantum protograph $A_Z$ of the $[[416,18,d\leq 20]]$ code in the limit of infinite $X$-bias. Here we see that the code has decoupled to a set of independent copies of the seed protograph $A$ and its transpose $A^T$. Both $A$ and $A^T$ have distance $26$ when converted into binary form, giving an infinite-bias $X$-distance of $d_X=26$ for the $[[416,18,d\leq 20]]$ code. As the $X$-bias is increased, we therefore see a reduction in the word error rate until a level commensurate with the classical performance of the seed codes is reached. From this point on, the word error rate plateaus as the code performance cannot exceed that of the original seed codes. The reason we do not see a similar plateau for the XZZX toric code in Fig.~\ref{fig:xzzx_css_comparison} is that it has an infinite-bias $X$-distance of $d_X=240$. The bias at which this code will plateau is therefore well beyond the regime which can be realistically simulated.

\begin{figure}
    \centering
    \input{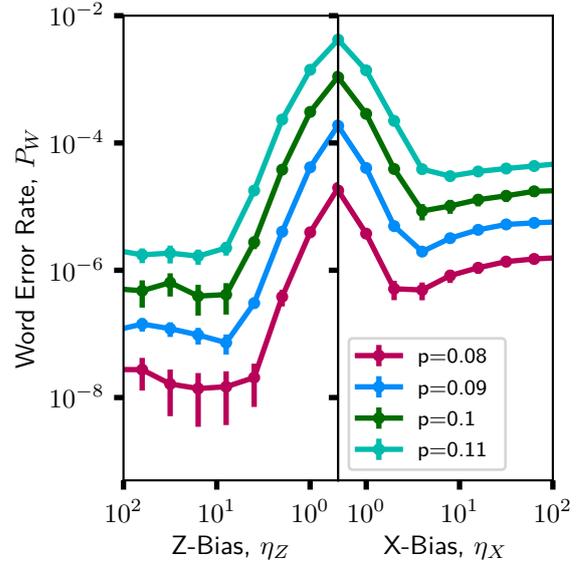}
    \caption{BP+OSD decoding of the $[[882,24,d\leq 24]]$ bias-tailored lifted product code for values of the physical error rate in the range $0.08\leq p \leq 0.11$. This plot is split into two parts: \textit{Left}. The decoding plot for $Z$-bias, with $\eta_Z$ increasing from right-to-left; \textit{Right}. The decoding plot for $X$-bias, with $\eta_X$ increasing from left-to-right. The two plots meet at the depolarising noise point where $\eta_X=\eta_Z=0.5$.}
    \label{fig:lpc2}
\end{figure}

The $[[416,18,d\leq20]]$ code is constructed from the bias-tailored lifted product of a pair of identical protographs. We therefore expect its performance to be the same for $X$- and $Z$-bias. In contrast, the $[[882,24,d\leq24]]$ code takes two different protographs as input (see Appendix \ref{app:protographs} for the exact form of the seed protographs). The decoding plot for the the bias-tailored $[[882,24,d\leq 24]]$ code is shown in Fig.~\ref{fig:lpc2} for increasing values of both the $X$- and $Z$-bias. For both bias-types, we again see an immediate reduction in the word error rate. However, the $X$-bias curves plateau at a word error rates approximately two orders of magnitude higher than the corresponding curves for $Z$-bias. This implies that of the two protographs, the one used to define the $X$-stabilisers corresponds to the better classical code.  

In Fig.~\ref{fig:lpc2}, we also note that the decoding curves of the $[[882,24,d\leq 24]]$ reach a minimum before levelling off at an increased value. This effect arises because of the specific form of the error model we are using. As the bias is increased, the error rate for the $H_Z$ component of the decoder increases whilst the error rate for $H_X$ component decreases. At a certain value of $\eta_X$, the error rate for the $H_X$ component is so low that the effective decoder failure rate is zero. As such, increasing $\eta_X$ further simply results in a larger error rate for $H_Z$, and therefore leads to an increase in the overall word error rate.

\subsection{Qubit overheads for quantum memories}

\begin{figure}
    \centering
    \input{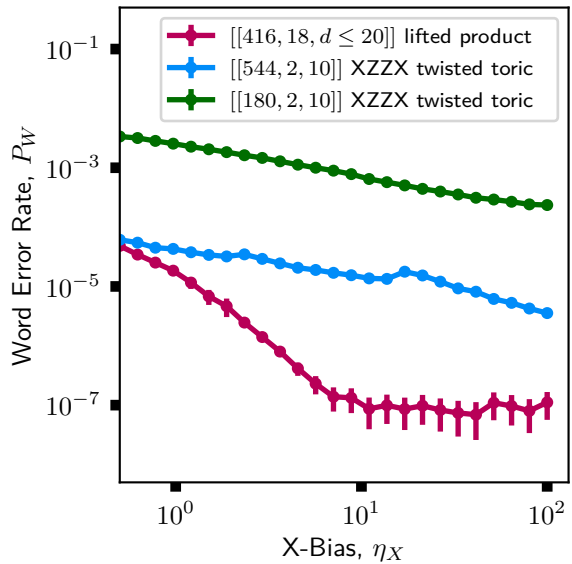}
    \caption{Decoding comparison between a bias-tailored lifted product code and two XZZX twisted toric codes. The toric codes are decoded using MWPM whilst the bias-tailored lifted code uses BP+OSD. For all simulations shown in this plot, the physical error rate is fixed at $p=0.06$. Error bars show three times the standard deviation.}
    \label{fig:overhead_comparison}
\end{figure}

The specific advantage of quantum LDPC codes over surface codes is their higher rate. Fig.~\ref{fig:overhead_comparison} shows the decoding plot for the $[[416,18,d\leq20]]$ bias-tailored lifted product code compared to two instances of the XZZX rotated toric code with parameters $[[180,2,10]]$ and $[[544,2,17]]$ respectively. Under depolarising noise, the simulated word error rate of the $[[416,18,d\leq20]]$ bias-tailored lifted product code is approximately equal to that of the $[[544,2,17]]$ XZZX twisted toric code. However, the $[[416,18,d\leq20]]$ code encodes $18$ logical qubits whilst the XZZX twisted toric code only encodes two. For the same word error rate, the quantum LDPC encoding therefore reduces the qubit overhead by a factor of nine compared to the toric code. Alternatively, a smaller toric code could be used so that the $18$ logical qubits can be encoded with reduced overhead. However, it is clear from decoding plot for the $[[180,2,10]]$ XZZX twisted toric code that this strategy imposes a severe penalty in terms of the word error rate. As the bias is increased, the word error rate of the $[[416,18,d\leq20]]$ code initially drops faster than the twisted toric code. Due to the plateau effect, we would expect the word rate of the lifted product code to eventually intersect with that of the $[[544,2,17]]$ twisted toric code. However, this would occur at an unrealistically high bias.   

\section{Summary \& outlook}

Physical qubits are typically subject to asymmetric noise where one species of error dominates over the other. As such, bias-tailoring provides a ripe opportunity for reducing the overhead of a QEC code. Prior to this work, studies conducted on the XZZX surface code have shown that it achieves remarkable performance in the high-bias regime. In this paper, we have extended these findings to show that bias-tailoring is also possible for the wider family of quantum LDPC codes.

\jr{
Our principal result is the introduction of the \textit{bias-tailored lifted product} as a method for constructing quantum codes for asymmetric noise channels. This is obtained via a modification of the stabilisers of the original lifted product defined in \cite{panteleev2020quantum}. We first showed that the XZZX toric code can be viewed as a special-case of the bias-tailored lifted product: both the stabiliser redefinition and boundary twists that characterise the XZZX toric code arise naturally from bias-tailored lifted product applied to two repetition codes. When applied to pairs of classical LDPC codes, the bias-tailored lifted product yields quantum LDPC codes with high-rate and good block-length-to-distance scaling. From this, we see that the bias-tailored lifted product provides a general framework for bias-tailoring, encapsulating both known examples of topological codes as well as quantum LDPC codes constructed using random methods.
}

We constructed examples of bias-tailored lifted product codes based on classical quasi-cyclic LDPC codes. Our Monte Carlo simulations of these codes under BP+OSD decoding showed that they outperform the toric codes for all realistic values of the bias. From this it is clear: connectivity-permitting, it is highly beneficial to consider a quantum LDPC encoding over the surface or toric code.

\joschkaedit{The performance of bias-tailored quantum LDPC codes derived from lifted products depends upon the quality of original seeds codes. This is due to fact that in the infinite-bias regime, the quantum code reduces to decoupled copies of the seed codes. As a result, it is important that the seed codes are optimised as much as possible. In this work, we have derived codes using the lifted product of quasi-cyclic codes. However, the lifted product can be applied to other families of codes. For instance, in \cite{panteleev21}, it is shown that quantum LDPC codes can be obtained using the lifted products of expander codes. It would be interesting to see whether such alternative constructions could yield codes at finite sizes (eg. $<1000$ qubits) that are competitive with quasi-cyclic lifted quantum LDPC codes.}

In this paper, we have considered a noise model were each qubit is subject to the same biased depolarising channel. In practice, however, each qubit on a quantum computer will be subject to a unique depolarising error channel: there is therefore the potential for the bias-tailoring to be more fine-grained. This idea has been explored for surface codes in \cite{dua22} and \cite{tiurev22}. In \cite{dua22}, random code deformations are investigated, whilst in \cite{tiurev22} a scheme for individually bias-tailoring each qubit is proposed. It would be interesting to see whether similar improvements in the performance of bias-tailored quantum LDPC codes can be achieved for non-identically distributed error models. Another interesting direction for future research would be to consider the bias tailoring of higher-dimensional topological codes, as has recently been explored for various families of codes defined on 3D lattices in \cite{huang22}.

All the simulations in this paper were carried out under the assumption of perfect syndrome readout. A natural follow up to this work is therefore to study code performance under a more realistic \textit{circuit-level} noise model. As part of this, a necessary consideration is that certain quantum circuit elements can transform one error type to another. For example, a two-qubit controlled-$Z$ gate propagates $X$- errors into $Z$-errors. Such propagation can negate the benefit of using a bias-tailored code. To counter this, it is necessary to engineer \textit{bias-preserving} two qubit gates. It has been proposed that the XZZX surface code could retain its high-bias performance via an implementation involving Kerr-cat qubits \cite{puri2021}. Similarly, bias-preserving gates will need to be engineered into the design of the non-local architectures used to implement quantum LDPC codes.

\joschkaedit{In the surface code, stabiliser extraction circuits are made fault tolerant by repeating measurements and decoding over time. A similar strategy could be pursued for quantum LDPC codes. However, this is likely to introduce a large decoding overhead due to the large the number of repetitions required for fault tolerance \cite{tansuwannont2022adaptive}. Additionally, decoding in this context must be carried out over an extensive space-time factor graph, a complication than can impact the benefits derived from bias-tailoring \cite{higgott2022fragile}.}

\joschkaedit{An alternative strategy for constructing fault tolerant quantum error correction codes is to use a so-called \textit{single-shot code}. Single-shot codes are designed to have redundant stabilisers that are protected by a classical code \cite{bombin2015single}. This allows imperfect measurements to be corrected, and the quantum code to be decoded using only a single round of stabiliser measurements. In \cite{campbell2019theory}, it was established that single-shot codes can be constructed using higher-dimensional homological products. Building on this, it has been shown that single-shot quantum LDPC codes can be efficiently decoded using BP+OSD \cite{quintavalle2021,higgott2022improved}. An interesting avenue for future research would be to investigate the construction of higher dimensional lifted products codes and their single-shot properties. For instance, a bias-tailored quantum LDPC code could be obtained by taking the lifted 3D hypergraph product over the ring of circulants. Such a code would be single-shot for one species of Pauli error. Extending this concept, fully single-shot codes could be obtained from lifted variants of the 4D hypergraph product.}

\joschkaedit{The BP+OSD decoder serves as a general-purpose decoder for quantum LDPC codes. While the belief propagation component operates in $O(N)$ time, where $N$ represents the block-length of the code, the ordered statistics component necessitates Gaussian elimination with an $O(N^3)$ runtime. Depending on the clock cycle of the selected qubit architecture, the OSD post-processing might present an obstacle to real-time decoding of quantum LDPC codes \cite{Raveendran21}. As a result, exploring alternative decoding strategies is crucial. One possibility is that the OSD decoding stage could be replaced by an alternative post-processing decoder such as union find \cite{Delfosse22,berent23}, small set-flip \cite{grospellier2020combining} or probabilistic flip \cite{scruby2023local}. Another strategy is to directly optimise the BP algorithm itself, for example through the use of machine learning, to derive improved update rules \cite{Liu_2019}. Finally, there are BP-variants that are specifically designed to mitigate the effects of trapping sets in quantum codes. These include generalised belief propagation \cite{old2022generalized}, stabiliser inactivation \cite{Crest2022StabilizerIF} and memory belief propagation \cite{Kuo_2022}. Such methods could reduce, or even remove, the need for resource intensive post-processing.}



It has long been known, both in theory and practice, that classical codes can (near) saturate the Shannon limit. Recently, it has been proved that \textit{optimal} families of lifted product codes exist with linear distance to block-length scaling \cite{panteleev21}. An interesting avenue for future work would be to study the numerical performance of such codes under different bias regimes. In particular, this could cast light on the ultimate capacity of quantum channels beyond the Hashing bound.

\section{Acknowledgements}
\subsection{People}
The authors thank Nithin Raveendran and Ben Brown for useful discussions. JR thanks Peter-Jan Derks, Shozab Qasim and Alex Townsend-Teague for comments and checking the manuscript. 

\subsection{Funding}
JR is funded by BMBF (RealistiQ) and the DFG (CRC 183). In this initial phase of this project, JR and ETC were  supported by the QCDA project (EP/R043825/1) which has received funding from the QuantERA ERA-NET Cofund in Quantum Technologies implemented within the European Union’s Horizon 2020 Programme. ETC's and AOQ's technical contributions were made while at the University of Sheffield. LZC acknowledges support from the Australian Research Council via the Centre of Excellence in Engineered Quantum Systems (EQUS) project number CE170100009.

\subsection{HPC resources}
The authors thank the HPC Service of ZEDAT, Freie Universität Berlin, for computing time \cite{fu_hpc}.

\subsection{Open source software}
The following open source software packages were used in this project: Numpy \cite{numpy}; Matplotlib \cite{matplotlib}; Scipy \cite{scipy}; LDPC \cite{roffe_ldpc}; BPOSD \cite{bp_osd}; pyMatching \cite{higgott2021pymatching}; Software for LDPC codes \cite{radford_neal}.

\subsection{The planet}
The carbon footprint associated with the numerical simulations in this paper are summarised below.
\begin{table}[H]
    \centering
\resizebox{!}{!}{%
\begin{tabular}[b]{l c}
\hline
\textbf{Numerical simulations} & \\
\hline
Total Kernel Hours [$\mathrm{h}$]& 77238\\
Thermal Design Power Per Kernel [$\mathrm{W}$]& 5.75\\
Total Energy Consumption [$\mathrm{kWh}$] & 444\\
Average Emission Of CO$_2$ [$\mathrm{kg/kWh}$]& 0.31\\
\hline
Total CO$_2$-Emission [$\mathrm{kg}$] & 138\\
\hline
\hline
\end{tabular}}
    \label{tab:my_label}
\end{table}
\noindent For context, $138$kg is the amount of carbon that would be emitted by the average German car over a journey of approximately $1000$km. The $\text{CO}_2$ emission per kWh used in the above table was the average value in Germany during 2020 according to European Environment Agency. The $138$kg emitted will be offset by donating to Atmosfair (\url{https://atmosfair.de}), a not-for-profit that promotes, develops and finances renewable energies in over fifteen countries worldwide. Guidance on the reporting the carbon cost of scientific research can be found in \cite{scientific_conduct}. 

\bibliography{main.bib}
\bibliographystyle{aps_new}

\appendix

\section{Parity check matrices}\label{app:parity_check_matrices}

In Example~\ref{example:hgp1}, a hypergraph product code is constructed from seed codes with the parameters $[16,4,6]$. The parity check matrix $H$ for these seed codes is given below:

\begin{equation}\label{eq:mkmn}
\arraycolsep=1pt\def\arraystretch{0.15}
\thickmuskip=0.25\thickmuskip
H=\left[\begin{array}{@{} cccccccccccccccc @{}}
    1&1&0&0&1&1&0&0&0&0&0&0&0&0&0&0\\ 
    0&0&1&0&0&0&1&1&0&0&0&0&1&0&0&0\\ 
    0&0&0&1&1&0&1&0&0&0&0&0&0&1&0&0\\ 
    0&0&0&0&0&1&0&0&1&1&0&0&0&0&0&1\\ 
    0&1&0&0&0&0&0&1&1&0&0&1&0&0&0&0\\ 
    0&0&0&0&0&0&0&0&1&0&0&0&1&1&1&0\\ 
    1&0&0&0&0&0&0&1&0&0&0&0&0&1&0&1\\ 
    0&0&0&1&0&1&0&0&0&0&1&0&1&0&0&0\\ 
    0&0&1&1&0&0&0&0&0&0&0&1&0&0&0&1\\ 
    0&0&0&0&1&0&0&0&0&1&1&1&0&0&0&0\\ 
    0&1&0&0&0&0&1&0&0&0&1&0&0&0&1&0\\ 
    1&0&1&0&0&0&0&0&0&1&0&0&0&0&1&0
    \end{array}
    \right]\rm.
\end{equation}

\section{Protographs} \label{app:protographs}

The $[[882,24,d\leq24]]$ code is constructed from the bias-tailored lifted product with $L=63$ of the following seed protographs  

\begin{align}
&A_1=\begin{bmatrix}
\lambda^{0}_L+\lambda^{1}_L+\lambda^{6}_L
\end{bmatrix}\\
&A_2=\begin{bmatrix}
\lambda^{36}_L&0&0&0&0&\lambda^{0}_L&\lambda^{9}_L\\ 
\lambda^{9}_L&\lambda^{36}_L&0&0&0&0&\lambda^{0}_L\\ 
\lambda^{0}_L&\lambda^{9}_L&\lambda^{36}_L&0&0&0&0\\ 
0&\lambda^{0}_L&\lambda^{9}_L&\lambda^{36}_L&0&0&0\\ 
0&0&\lambda^{0}_L&\lambda^{9}_L&\lambda^{36}_L&0&0\\ 
0&0&0&\lambda^{0}_L&\lambda^{9}_L&\lambda^{36}_L&0\\ 
0&0&0&0&\lambda^{0}_L&\lambda^{9}_L&\lambda^{36}_L
\end{bmatrix}\rm.
\end{align}
The upper-bound on the minimum distance is an estimate based on the lowest-weight logical operator observed during simulation with the BP+OSD decoder. This code is equivalent to the CSS \textit{generalised hypergraph product} code of the same size first presented in \cite{Panteleev_2019}.

\section{The Hashing bound under biased-noise}\label{app:hashing}

Given a noisy channel, what is the maximum amount of information that can be reliably transferred using error correction? In the classical setting, the answer to this question is given by the Shannon noisy coding theorem \cite{shannon49}. For a given error channel, the Shannon noisy coding theorem tells us that \textit{good} error correction codes exist with the ability to arbitrarily suppress the logical error rate. However, the rate $r = k/n$ of such good codes must fall below a certain value known as the Shannon limit. To illustrate, we consider the simple case of the classical binary symmetric error channel where each bit is independently subject to a bit-flip with probability $p_X$. For this channel, the Shannon limit $r_S$ is given by
\begin{equation}\label{eq:shannon_bound}
r_S=1-\mathcal{H}(p_X)\rm,
\end{equation}
where $\mathcal{H}(p_X)$ is the binary entropy defined as
\begin{equation}
    \mathcal{H}(p_X)=-(1-p_X)\log_2(1-p_X) -p_X \log_2(p_X)\rm.
\end{equation}
The Shannon limit implies that \textit{perfect} error correction is in principle possible with error correction codes with code rate $r \leq r_S$, whilst it is impossible with code rate $r > r_S$. Various classical error correction codes -- including LDPC, turbo, and polar codes -- are known to be \textit{optimal} in the sense that they operate at rates approaching the Shannon limit whilst being efficiently decodable~\cite{gallager1962, mackay1997near, berrou1996near, arikan2009channel}.

The Shannon noisy coding theorem can be directly extended to quantum channels. This provides a Shannon limit for stabiliser codes referred to as the Hashing bound. For the noisy quantum channel $\mathcal{E}_Q(\rho)$ defined in Eq.~(\ref{eq:error_channel}) the Hashing bound $r_H$ is given by~\cite{bennett1996mixed}
\begin{equation}\label{eq:hashing_bound}
    r_H=1-\mathcal{H}_Q(p)\rm.
\end{equation}
Here, $\mathcal{H}_Q(p)$ is the Pauli entropy function defined as
\begin{equation}
\begin{split}
    \mathcal{H}_Q(&p)=-(1-p)\log_2(1-p)\\-&p_X \log_2(p_X)-p_Y \log_2(p_Y)-p_Z \log_2(p_Z)\rm,
\end{split}
\end{equation}
where $p$ is the total physical error rate $p=p_X+p_Y+p_Z$. \jr{It should be noted that the Hashing bound is derived using methods from classical information theory: quantum effects such as entanglement and degeneracy, which increase the actual capacity of the quantum channel~\cite{divincenzo1998quantum}, are not accounted for. As such, the Hashing bound should be treated as a lower-limit on the achievable rate of reliable information transfer over a quantum channel.} In fact, in certain settings, stabiliser codes have been shown to achieve logical error rate suppression beyond \etc{the Hashing bound \cite{shor1996quantum,ataides2020}}. Quantifying the true capacity of a quantum channel remains an open problem.

\joschka{Given an existing error correction code with a known rate it is possible to calculate its operational regime by calculating the Hashing probability $p_H(r)$. The Hashing probability is the value obtained for the total physical probability $p$ when $r_H$ in Eq.~(\ref{eq:hashing_bound}) is set to a fixed value $r_H=r$\etc{, which is readily obtained using a numerical solver.} The Hashing probability $p_H(r)$ tells us the physical error rate below which it is possible to arbitrarily suppress the logical error rate.}  

The toric code and its topological cousins are often benchmarked in terms of the code threshold $p_{TH}$. Provided the physical error rate of the qubits is below the threshold $p < p_{TH}$, increasing the size of a code will reduce the logical error rate $p_L$. In the sub-threshold regime, and in the limit of large block lengths where the code rate $r\rightarrow0$, we expect the logical error rate to be arbitrarily suppressed $\lim_{r\rightarrow0 } p_L = 0$. For these codes, an estimate of the code threshold can therefore be derived from the \textit{zero-rate} Hashing probability $p_H(0)$.

\begin{figure}
    \centering
    \input{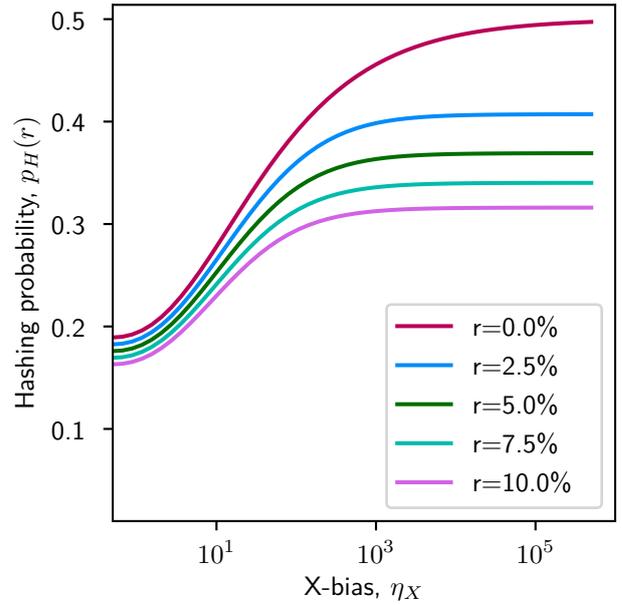}
    \caption{Plots of of the hashing probability against the $X$-bias for code rates between $R=10\%$ and $R=0\%$. The $x$-axis intersects the $y$-axis at $\eta_X=0.5$ which corresponds to depolarising noise. The Hashing probabilities were obtained by numerically solving Eq.~(\ref{eq:hashing_bound}) for $p$ for fixed values of the rate $r_H=r$.}
    \label{fig:hashing_probs}
\end{figure}

We now explore how adjusting the qubit bias affects the Hashing limit of the quantum channel. Fig.~\ref{fig:hashing_probs} shows the Hashing probability plotted against the $X$-bias for code error rates between $R=10\%$ and $R=0$. Here the $X$-bias is increased setting $p_Y = p_Z$ and raising the value of $p_X$ whilst keeping the total physical error rate fixed $p = p_X + p_Y + p_Z$. The red line shows the zero-rate Hashing probability and therefore gives an estimate of the code threshold. At the origin, which corresponds to depolarising noise where $p_X = p_Y = p_Z$, the zero-rate Hashing limit is $18.9\%$. As the bias grows, the Hashing limit rises until converging to a value of $50\%$. In the infinite-bias regime, it is clear that the Hashing bound defined in Eq.~(\ref{eq:hashing_bound}) is equivalent to the classical Shannon limit in Eq.~(\ref{eq:shannon_bound}).

Analysing the quantum noise model in terms of the Hashing bound implies that there is much to be gained from accounting for noise asymmetry in the design of quantum codes. \jr{Up to this point, we have explored the zero-rate Hashing bound that only applies to codes, such as the toric code, which have vanishing rate in the asymptomatic limit. In addition to the zero-rate line, Fig.~\ref{fig:hashing_probs} also plots the Hashing probability for various finite values of the rate. For all rates, we see an improvement in the Hashing capacity as the bias is increased. As such, we also expect bias-tailoring to be useful for codes, such as quantum LDPC codes, that have finite rate in the asymptotic limit}.

\section{Simulation algorithm}\label{app:simulation}
In this Appendix we describe implementation details of our Monte Carlo simulations of quantum LDPC decoding. We present all algorithms in pseudo-code. The actual simulation code can be found in the external repository for this project \cite{biased_qldpc_code}.

\subsection{Two-stage decoding of Hadamard rotated codes}
For CSS codes, it is possible to implement a two-stage decoding procedure where the $X$- and $Z$-stabilisers are treated as separate codes. The stabilisers of the bias-tailored codes explored this work differ from their \textit{standard} CSS code counterparts by a Hadamard rotation on the sector-two qubits. The resultant codes are non-CSS which would ordinarily necessitate having to decode the entire quantum parity check matrix in a single stage. Fortunately, for the specific case of the Hadamard rotated codes, a standard CSS two-stage decoding routine remains possible. To illustrate, we first consider the effect the Hadmard rotation has on the syndrome equations. We take the case of a standard CSS code obtained from a product construction (eg. hypegraph product, lifted product etc.) such that its qubits can be separated into sector-one and sector-two qubits. In this setting, the quantum parity check matrix has the form
\begin{equation*}
	H_{\rm CSS}=\left[\begin{array}{@{} c|c @{}}
		0&H'_Z\\
		H'_X&0
	\end{array}\right]=\left[\begin{array}{@{} cc|cc @{}}
		0&0&H_{Z1} & H_{Z2}\\
		H_{X1}&H_{X2}&0&0
	\end{array}\right]\rm.
\end{equation*} 
Following a Hadmard rotation, the above parity check matrix is transformed to 
\begin{equation*}
	H_{\rm HR}=\left[\begin{array}{@{} cc|cc @{}}
		0&H_{Z2}&H_{Z1} & 0\\
		H_{X1}&0&0&H_{X2}
	\end{array}\right]\rm.
\end{equation*} 
To compute the syndrome we apply the Hadamard rotated stabilisers $H_{HR}$ to the error vector $\mathbf{e}=(\mathbf{e_X}|\mathbf{e_Z})$, where the $X$- and $Z$-components are split into sector-one and sector-two qubits such that $\textbf{e}_X=\left(\textbf{e}_{X1},\textbf{e}_{X2}\right)$ and $\textbf{e}_Z=\left(\textbf{e}_{Z1},\textbf{e}_{Z2}\right)$. The syndrome equation then reads
\begin{equation*}
\begin{split}
   (\mathbf{s}_X&|\mathbf{s}_Z)= H_{\rm HR}\star\mathbf{e}=\\&(H_{Z1}\cdot \mathbf{e}_{X1}+H_{Z2}\cdot\mathbf{e}_{Z2}|H_{X1}\cdot\mathbf{e}_{Z1}+H_{X2}\cdot\mathbf{e}_{X2})\rm.
 \end{split}
\end{equation*}
It is now clear that decoding for the Hadamard rotated codes decouples into two problems
\begin{equation}\label{eq:hr_syndromes}
\begin{split}
\textbf{s}_X=H'_Z\cdot\left(\textbf{e}_{X1},\textbf{e}_{Z2}\right)\rm,\\
\textbf{s}_Z=H'_X\cdot \left(\textbf{e}_{Z1},\textbf{e}_{X2}\right)\rm,
\end{split}
\end{equation}
where $H'_X$ and $H'_Z$ are the CSS parity check matrices prior to the Hadamard rotation. From this, we see that for the purposes of simulation it is sufficient to simply Hadamard rotate the error vector whilst retaining the stabilisers in their original CSS form. In practice, this is achieved by modifying the error model that is input to the simulation algorithm.

\subsection{Main decoding simulation}

Algorithm~\ref{alg:monte_carlo} gives a pseudocode description of the Monte Carlo simulation used to access the performance of the quantum LDPC codes in this paper. Lines~\ref{line:start_hr}-\ref{line:end_hr} show how the Hadmard rotation is applied by modifying the original error model so that the $N-N_1$ qubits in sector-two have $p_X$ swapped with $p_Z$. This ensures that the randomly generated error vectors have the Hadamard rotated form from Eq~(\ref{eq:hr_syndromes}). In our simulations one of two decoding methods are used: 1) minimum-weight perfect matching (MWPM); 2) belief propagation plus ordered statistics decoding (BP+OSD). These decoders are described qualitatively in Section~\ref{sec:decoders}. Both decoding methods act over the factor graph of the code and accept prior information about the error model.

\begin{breakablealgorithm}
	\caption{\textsc{Simulation: decoding bias-tailored codes}}
	\label{alg:monte_carlo}
	\begin{algorithmic}[1]
		\Require Decoder; CSS parity check matrices $H'_X$ and $H'_Z$; CSS logical operator matrices $L'_X$ and $L'_Z$; Code block length N; Number of qubits in sector-one $N_1$; Qubit error rates for Pauli errors $\{p_X,p_Y,p_Z\}$.
		\Ensure Decoding Success or Failure
		\State $ \mathbf{e}_X \gets \mathbf{0}$ \Comment{Qubit $X$-error}
		\State $\mathbf{e}_Z \gets \mathbf{0}$ \Comment{Qubit $Z$-error}
		\State $\mathbf{p}_X \gets \mathbf{0}$ \Comment{Priors for $ \mathbf{e}_X$}
		\State $\mathbf{p}_Z \gets \mathbf{0}$  \Comment{Priors for $\mathbf{e}_Z$}

		\LineComment{Hadamard rotate error model}
		\For{$j\gets 1$ \textbf{to} $N_1$} 
		\State $\mathbf{p}_X[j] \gets p_X+p_Y$ \Comment{$X$-priors on sector-1}
		\State $\mathbf{p}_Z[j] \gets p_Z+p_Y$	\Comment{$Z$-prior on sector-1}
		\EndFor
        \For{$j\gets N_1+1$ \textbf{to} $N$}\label{line:start_hr}
		\State $\mathbf{p}_X[j] \gets p_Z+p_Y$ \Comment{$X$-priors on sector-2}
		\State $\mathbf{p}_Z[j] \gets p_X+p_Y$	\Comment{$Z$-prior on sector-2}
        \EndFor \label{line:end_hr}

		\LineComment{Generate random qubit error}
		\For{$j\gets 1$ \textbf{to} $N$}
		\State $\zeta\gets$ random number between $0$ and $1$
		\If{$\zeta<\textbf{p}_X[j]$} 
		\State $\mathbf{e}_X[j]\gets 1$
		\State $\mathbf{e}_X[j]\gets 0$
		\ElsIf{$\textbf{p}_X[j]<\zeta<\textbf{p}_X[j]+\textbf{p}_Z[j]$}
		\State $\mathbf{e}_X[j]\gets 0$
		\State $\mathbf{e}_Z[j]\gets 1$
		
		\ElsIf{$\textbf{p}_X[j]+\textbf{p}_Z[j]<\zeta<\textbf{p}_X[j]+\textbf{p}_Z[j]+\textbf{p}_Y[j]$}
		\State $\mathbf{e}_X[j]\gets 1$
		\State $\mathbf{e}_Z[j]\gets 1$
		\EndIf
		
		\EndFor

		\LineComment{Calculate syndromes}
		
		\State $\mathbf{s}_X=H'_Z\cdot \mathbf{e}_X$ \Comment{$X$-syndrome}
		\State $\mathbf{s}_Z=H'_X\cdot \mathbf{e}_Z$ \Comment{$Z$-syndrome}

		\LineComment{$X$-error decoding}
		\State Use the decoder with priors $\mathbf{p}_{X}$ to obtain recovery $\mathbf{r}_{X}$ s.t.\ $H'_Z\cdot \mathbf{r}_{X}=\mathbf{s}_{X}$
		
		\LineComment{Channel update subroutine. The priors $\mathbf{p}_Z$ for the second round of decoding can be updated based on the output of the first round of decoding. This helps account for correlations between the $X$- and $Z$-decoding rounds that arise due to independent $Y$-errors. See Algorithm~\ref{alg:channel_update} for pseudocode of this subroutine.}
		
		\LineComment{$Z$-error decoding}
		\State Use the decoder with priors $\mathbf{p}_{Z}$ to obtain recovery $\mathbf{r}_{Z}$ s.t.\ $H'_X\cdot \mathbf{r}_{Z}=\mathbf{s}_{Z}$

		\LineComment{Check for logical errors}
		\State $\mathbf{r'}_{X}=\mathbf{r}_{X}+\mathbf{e}_{X}$ \Comment{Residual $X$-error}
		\State $\mathbf{r'}_{Z}=\mathbf{r}_{Z}+\mathbf{e}_{Z}$ \Comment{Residual $Z$-error}
		
		\If{$L'_X\cdot \mathbf{r'}_{Z}\neq\textbf{0}$}
		\State \Return Failure
		\ElsIf{$L'_Z\cdot \mathbf{r'}_{X}\neq0$}
		\State \Return Failure
		\Else
		\State \Return Success
		\EndIf
	\end{algorithmic}
\end{breakablealgorithm}

\subsection{Accounting for correlated errors}
\begin{figure}
    \centering
    \input{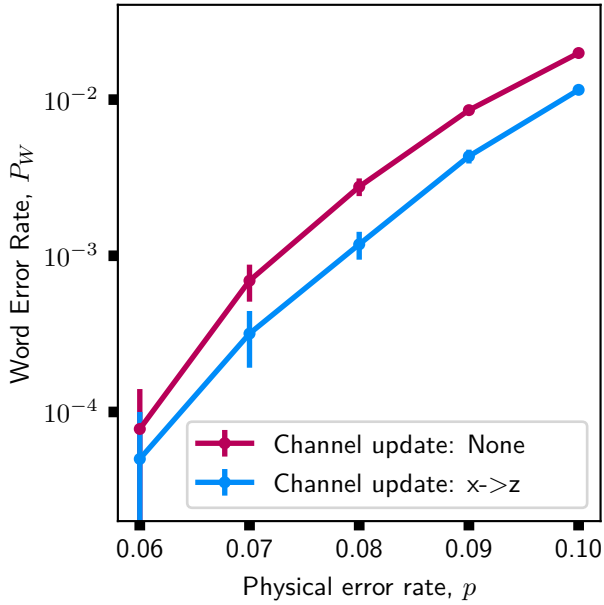}
    \caption{BP+OSD decoding of the $[[416,18,d\leq20]]$ quantum LDPC code with and without the channel update sub-routine described in Algorithm \ref{alg:channel_update}. }
    \label{fig:channel_update_comparison}
\end{figure}
Pauli-Y errors introduce correlations between the $\textbf{e}_X$ and $\textbf{e}_Z$ error vectors. Failing to account for these correlations can reduce the logical error rate of the decoding. Consider the scenario, as in Algorithm \ref{alg:monte_carlo}, where the $X$-errors are decoded before the $Z$-errors. Prior to the first round of decoding, we assign the probability that bit-$i$ is set to one in the error vector $\textbf{e}_X$ to be $\textbf{p}_X[i]=p_X+p_Y$: i.e. this is the combined probability that either an $X$- or a $Y$-error occurs. Now, imagine that the decoder tells us that the bit-$i$ in $\textbf{e}_X$ is zero $\textbf{e}_X[i]=0$. In the second round of decoding, the probability that bit-$i$ is $\textbf{e}_Z[i]=1$ is now given by
\begin{equation}\label{eq:channel_update_1}
    \textbf{p}_Z[i]=\frac{p_Z}{1-p_X-p_Y}\rm.
\end{equation}
In the above, probability of $\textbf{p}_Z[i]$ is adjusted to account for the fact that that $Y$-errors are no longer possible. Similarly, if the outcome of the fist round of decoding is $\mathbf{e}_X[i]=1$ then the the corresponding probability $\mathbf{p}_Z[i]$ is updated to
\begin{equation}\label{eq:channel_update_2}
    \textbf{p}_Z[i]=\frac{p_Y}{p_X+p_Y}\rm.
\end{equation}
These channel updates are implemented via the subroutine described in pseudocode below
\begin{breakablealgorithm}
	\caption{\textsc{Channel update subroutine}}
	\label{alg:channel_update}
	\begin{algorithmic}[1]
		\Require Stage 1 decoding output $\textbf{r}_X$; Code block length N; Qubit error rates for Pauli errors $\{p_X,p_Y,p_Z\}$.
		\Ensure Updated $Z$-priors $\textbf{p}_Z$
		\State $\mathbf{p}_Z \gets \mathbf{0}$  \Comment{Priors for $\mathbf{e}_Z$}
		
		\For{$j\gets 1$ \textbf{to} $N$} 
		\If{$\textbf{r}_X[i]=0$}
		\State $\textbf{p}_Z[i]=p_Z/(1-p_X-p_Y)$
		\ElsIf{$\textbf{r}_X[i]=1$}
		\State $\textbf{p}_Z[i]=p_Y/(p_X+p_Y)$
		\EndIf
	    \EndFor
		\State \Return $\textbf{p}_Z$
	\end{algorithmic}
\end{breakablealgorithm}
In this section, we have described channel updates from $X$->$Z$. However, updates in the reverse direction can easily be derived by swapping the roles of $p_X$ with $p_Z$ in Eqs.~(\ref{eq:channel_update_1}-\ref{eq:channel_update_2}). For the simulations in this work, we implemented channel updates for the codes decoded using BP+OSD. This was possible, as the channel information for BP+OSD can be updated on-the-fly without affecting the runtime. We did not implement channel updates with the MWPM decoder, as updating the channel information for this decoder necessitates re-running a pre-processing routine (Dijkstra's algorithm) that compiles a matching database.

Fig.~\ref{fig:channel_update_comparison} shows the word error rate against the physical depolarising error rate for the $[[416,18,d\leq20]]$ quantum LDPC code decoded using BP+OSD. The red line shows the case when no channel update is performed, whilst for the blue line a channel update from $X$->$Z$ performed. From this it is clear that the channel update subroutine improves the word error rate for all values of the physical error rate.

\end{document}